

\newif\iffigs\figsfalse
\figstrue

\newif\ifamsf\amsffalse
\amsftrue

\newif\ifscrf\scrffalse
\scrftrue

\input harvmac
\iffigs
  \input epsf
\else
  \message{No figures will be included. See TeX file for more
information.}
\fi
\catcode`\@=11
\newbox\Bbox
\newif\ifBinsert
\newdimen\fullvsize \fullvsize=\vsize
\newtoks\dfoutput \dfoutput=\the\output
\ifx\answ\bigans\else
\output={
  \ifBinsert
    \if L\l@r
       \the\dfoutput
    \else
       \count1=2
       \message{Doing wide insert.}
       \shipout\vbox{\speclscape
          {\hsize=\fullhsize \makeheadline}
          \hbox to\fullhsize{\hfil\box\Bbox\hfil}
          \hbox to\fullhsize{\box\leftpage\hfil
                \leftline{\vbox{\pagebody\makefootline}}}}
       \global\let\l@r=L
       \advancepageno
       \global\Binsertfalse
       \global\vsize=\fullvsize
    \fi
  \else
    \if B\botmark
      \if R\l@r
         \mark{}
         \global\Binserttrue
	 \the\dfoutput
         \global\advance\vsize by -\ht\Bbox
      \else
         \the\dfoutput
      \fi
    \else
      \the\dfoutput
    \fi
  \fi
}
\fi
\def\Title#1#2{\nopagenumbers\abstractfont\hsize=\hstitle\rightline{#1}%
\vskip 0.5in\centerline{\titlefont #2}\abstractfont\vskip .5in\pageno=0}
\Title{\vbox{\baselineskip12pt\hbox{IASSNS-HEP-93/38}\hbox{CLNS-93/1236}}}
{\vbox{\centerline{ Calabi-Yau Moduli Space, Mirror Manifolds and }
    \vskip2pt\centerline{Spacetime Topology Change in String Theory}}}

\centerline{Paul S. Aspinwall,%
\footnote{${}^\dagger$}{
School of Natural Sciences, Institute for Advanced Study,
Princeton, NJ \ 08540.
}
Brian R. Greene%
\footnote{${}^\sharp$}{
F.R. Newman Laboratory of Nuclear Studies,
Cornell University, Ithaca, NY \ 14853.
}
and David R. Morrison%
\footnote{${}^*$}{
School of Mathematics, Institute for Advanced Study,
Princeton, NJ \ 08540.
On leave from:  Department of Mathematics, Duke University,
Box 90320, Durham, NC \ 27708.
}}

\vskip .3in

\vskip .3in

\noblackbox

\font\eightrm=cmr8 \font\eighti=cmmi8
\font\eightsy=cmsy8 \font\eightbf=cmbx8
\font\eightit=cmti8 \font\eightsl=cmsl8 \skewchar\eighti='177
\skewchar\eightsy='60

\def\eightpoint{\def\rm{\fam0\eightrm}
\textfont0=\eightrm \scriptfont0=\fiverm \scriptscriptfont0=\fiverm
\textfont1=\eighti \scriptfont1=\fivei \scriptscriptfont1=\fivei
\textfont2=\eightsy \scriptfont2=\fivesy \scriptscriptfont2=\fivesy
\textfont\itfam=\eighti
\def\it{\fam\itfam\eightit}\def\sl{\fam\slfam\eightsl}%
\textfont\bffam=\eightbf \def\bf{\fam\bffam\eightbf}\rm}

\font\bigrm=cmr10 scaled \magstephalf
\def\inbar{\,\vrule height1.5ex width.4pt depth0pt}
\font\cmss=cmss10 \font\cmsss=cmss8 at 8pt
\def\BZ{\relax\ifmmode\mathchoice
{\hbox{\cmss Z\kern-.4em Z}}{\hbox{\cmss Z\kern-.4em Z}}
{\lower.9pt\hbox{\cmsss Z\kern-.36em Z}}
{\lower1.2pt\hbox{\cmsss Z\kern-.36em Z}}\else{\cmss Z\kern-.4em Z}\fi}
\def\IC{\relax\hbox{$\inbar\kern-.3em{\rm C}$}}
\def\IP{\relax{\rm I\kern-.18em P}}
\def\IQ{\relax\hbox{$\inbar\kern-.3em{\rm Q}$}}
\def\IR{\relax{\rm I\kern-.18em R}}
\def\AAA{{\cal A}}
\def\cpR{{\cal R}}
\def\MM{{\cal M}}
\def\SS{{\cal S}}
\newfam\bblfam
\newfam\gothfam
\newfam\scrfam
\ifamsf
\font\tenbbl=msbm10
\font\eightbbl=msbm8
\font\tengoth=eufm10
\font\eightgoth=eufm8
\ifscrf
\font\tenscr=rsfs10
\font\eightscr=rsfs7
\fi
\def\tenpoint{\def\rm{\fam0\tenrm}
\textfont0=\tenrm \scriptfont0=\sevenrm \scriptscriptfont0=\fiverm
\textfont1=\teni  \scriptfont1=\seveni  \scriptscriptfont1=\fivei
\textfont2=\tensy \scriptfont2=\sevensy \scriptscriptfont2=\fivesy
\textfont\itfam=\tenit \def\it{\fam\itfam\tenit}
\def\footnotefont{\ninepoint}
\textfont\bffam=\tenbf
\def\bf{\fam\bffam\tenbf}\def\sl{\fam\slfam\tensl}
\textfont\bblfam=\tenbbl \scriptfont\bblfam=\eightbbl
\scriptscriptfont\bblfam=\eightbbl\def\bbl{\fam\bblfam\tenbbl}
\textfont\gothfam=\tengoth \scriptfont\gothfam=\eightgoth
\scriptscriptfont\gothfam=\eightgoth\def\goth{\fam\gothfam\tengoth}
\ifscrf\textfont\scrfam=\tenscr \scriptfont\scrfam=\eightscr
\scriptscriptfont\scrfam=\eightscr\def\scr{\fam\scrfam\tenscr}\fi\rm}
\def\BZ{{\bbl Z}}
\def\IC{{\bbl C\hskip0.5pt}}
\def\IP{{\bbl P}}
\def\IQ{{\bbl Q}}
\def\IR{{\bbl R}}
\ifscrf
\def\AAA{{\scr A}}
\def\cpR{{\scr R}}
\def\MM{{\scr M}}
\def\SS{{\scr S}}
\fi
\fi  

\def\Tr#1{\hbox{{\bigrm Tr}\kern-1.05em \lower2.1ex \hbox{$\scriptstyle#1$}}\,}

\def\ex#1{\hbox{$\> e^{#1}\>$}}

\def\WCP#1#2{\hbox{$\hbox{\IP}^{#1}_{\{#2\}}$}}

\def \LG{Landau-Ginzburg}

\def\CP#1{{\IP}^{#1}}

\def\tilde{\widetilde}

\def\Hom{\mathop{\rm Hom}}
\def\Spec{\mathop{\rm Spec}}

\def\X{X}
\def\W{Y}

\def\CY{Calabi-Yau}
\def\to{\rightarrow}
\def\tto{\longrightarrow}
\def\AA{{\AAA}}
\def\AAO{{\Xi}}
\def\Psf{\Sigma^\prime}
\def\tablerule{\noalign{\hrule}}

\def\vol{\mathop{\rm vol}}
\def\spn{\mathop{\rm span}}

We analyze the moduli spaces of \CY\ threefolds and
their associated conformally invariant nonlinear $\sigma$-models and
show that they are described by an unexpectedly rich geometrical structure.
Specifically,
the K\"ahler sector of the  moduli space  of such \CY\ conformal theories
admits a decomposition into  adjacent domains  some of which
correspond to the (complexified) K\"ahler cones of  topologically
distinct manifolds. These domains  are
separated by walls corresponding to singular \CY\ spaces in which
the spacetime
metric has degenerated in certain regions.
We show that the union of these
domains is isomorphic to the complex
structure moduli space of a {\it single\/} topological Calabi-Yau space ---
the mirror. In this way we resolve a puzzle for mirror symmetry raised by the
apparent
{\it asymmetry\/} between the K\"ahler and complex structure
moduli spaces of a Calabi-Yau manifold. Furthermore,
using mirror symmetry,
we show that   we can
interpolate in a physically smooth manner between {\it any\/}
two theories represented by distinct points in
the K\"ahler moduli space, even if such  points correspond to
topologically distinct spaces.
Spacetime topology change in string theory,
therefore, is realized by the most basic operation of deformation by
a truly marginal operator. Finally,
this work also yields some important insights on
the nature of orbifolds in string theory.

 \Date{8/93}

\newsec{Introduction}

Over the years, research on string theory has followed two main paths.
One such path has been the attempt to extract detailed and specific low
energy models from string theory in an attempt to make contact with observable
physics. A wealth of research towards this end has shown conclusively that
string theory contains within it all of the ingredients essential to
building the standard model and that, if we are maximally optimistic about
those things we do not understand at present, fairly realistic
low energy models can be constructed.
The second research path
has focused on those properties of the theory  which are generic to all models
based on strings and which are difficult, if not impossible, to accommodate
in a theory based on point particles. Such properties single out
characteristic ``stringy''
phenomena and hence constitute the true distinguishing
features of string theory. With our present inability to extract definitive
low energy predictions from string theory, there is strong motivation to
study these generic features.

\nref\rDHVW{L. Dixon, J. Harvey, C. Vafa and E. Witten, Nucl. Phys.
{\bf B261} (1985) 678; Nucl. Phys. {\bf B274} (1986) 285.}
\nref\rDixon{L. Dixon, in {\it Superstrings, Unified Theories and
Cosmology 1987}\/ (G. Furlan et al., eds.), World Scientific, 1988, p. 67.}
\nref\rLVW{W. Lerche, C. Vafa and N. Warner, Nucl. Phys. {\bf B324}
(1989) 427.}
\nref\rCLS{P. Candelas, M. Lynker and R. Schimmrigk, Nucl. Phys. {\bf B341}
(1990) 383.}
\nref\rGP{B.R. Greene and M.R. Plesser, Nucl. Phys. {\bf B338} (1990) 15.}%

One such feature was identified  some time ago in the work
of \rDHVW. These authors showed that whereas point particle theories appear
to require a smooth background spacetime, string theory is well defined
in the presence of a certain class of spacetime singularities:
toroidal quotient singularities.
Another characteristically stringy
feature is that of mirror symmetry and mirror manifolds.
Mirror symmetry was conjectured
based upon naturality arguments in \refs{\rDixon,\rLVW},
was strongly suggested by the computer studies of \rCLS,
and was established to
exist in certain cases
by direct construction in \rGP.
The phenomenon of mirror manifolds shows that
vastly different spacetime backgrounds can give rise to identical
physics --- something quite unexpected in nonstring based theories
such as general relativity and Kaluza-Klein theory.
The present work is a natural progression along these lines of research.
By using mirror symmetry we show, amongst other
things, that the topology of spacetime can
change by passing through a mathematically singular space while
the physics of string theory is perfectly well behaved.

\nref\rMODULISPACE{P. Candelas and X.C. De La Ossa, Nucl. Phys. {\bf
B355} (1991) 455\semi
A.Strominger, Commun. Math. Phys. {\bf 133} (1990)
163\semi
S. Cecotti. S. Ferrara and L. Girardello, Phys. Lett. {\bf 213B}
(1989) 443.}

 From a more general vantage point, the present work focuses on the structure
of the moduli spaces of \CY\ manifolds and their associated superconformal
nonlinear $\sigma$-models.
There has been much work on this subject over the last
few years
\rMODULISPACE,
mainly focusing on local
properties. The burden of the sequel is to show that an investigation
of more global properties reveals a remarkably rich structure.
As we shall see, whereas previous studies \rMODULISPACE\ have, as a
prime example, considered the local geometry of the complexified K\"ahler
cone for a \CY\ space of a fixed topology, we find that a more
global perspective shows that numerous such K\"ahler cones for
topologically distinct \CY\ spaces (and other more exotic entities to
be discussed) fit together by adjoining along common walls to form
what we term
the {\it enlarged K\"ahler moduli space}. The common walls of these
K\"ahler cones correspond to  metrically degenerate \CY\ spaces
in which some homologically nontrivial cycle has zero volume.
In other words, points in these walls
correspond to a configuration in which some nontrivial subvariety
of the \CY\ space (or, in fact, the whole \CY\ itself!) is shrunk down
to a point. Nonetheless, we show, by making use of mirror symmetry,
that a generic point in such a wall
corresponds to a perfectly well behaved conformal field theory.
In fact, the generic point in such a wall
has {\it no special significance\/}
from the point of view of conformal field theory and hence none from the point
of view of physics. As is familiar from previous studies,
one can move around a path in moduli space by changing the expectation value
of truly marginal operators. We show that a typical path passes through
these walls without any unusual physical consequence. Hence, this makes
it evident that it is physically incomplete to study a single
K\"ahler cone.

There are four main implications of this newfound need to pass from
a single complexified K\"ahler cone to the enlarged K\"ahler moduli space:

\nref\rViehweg{E. Viehweg, Invent. Math. {\bf 101} (1990) 521.}

1) Mirror symmetry, combined with standard reasoning from conformal field
theory, leads one to the conclusion that if $X$ and $Y$ are a mirror pair
of \CY\ spaces then the complexified K\"ahler moduli space of $X$ is isomorphic
to the complex structure moduli space of $Y$ and vice versa \rGP. This is
a puzzling statement mathematically because a  complexified K\"ahler
moduli space
is a bounded domain (as we shall discuss this is due to the usual
constraints that the K\"ahler form should yield positive volumes) whereas
a complex structure moduli space is not bounded, but is rather of the
form $A-B$ with $A$ and $B$ subvarieties of some projective space \rViehweg.
The present work shows that the true
object which is mirror to  a complex structure moduli space is not a single
complexified K\"ahler cone, but rather the enlarged K\"ahler moduli
space introduced above. We show that the latter has an identical
mathematical description (using toric geometry) as the mirror's complex
structure moduli space, thus resolving this important issue.

\nref\rAGM{P.S. Aspinwall, B.R. Greene and D.R. Morrison, Phys. Lett.
{\bf 303B} (1993) 249.}
\nref\rTY{G. Tian and S.-T. Yau, in {\it Mathematical Aspects of String
Theory}\/ (S.-T. Yau, ed.), World Scientific, 1987, p. 543.}
\nref\rflops{J.~Koll\'ar,
  Nagoya Math. J. {\bf 113} (1989) 15.}

2) In the example considered in \rAGM\ and discussed in greater detail
here, the enlarged K\"ahler moduli space consists of 100 distinct regions.
In the more familiar example of the mirror of the quintic hypersurface,
we expect the number of regions in the enlarged K\"ahler moduli space
to be far
greater -- possibly many orders of magnitude greater -- than in the example
studied here.
An important question is to give the
physical interpretation of the theories in each region. In \rAGM\ we found
that five of the 100 regions  in our example
were interpretable as the complexified
K\"ahler moduli spaces of five topologically distinct \CY\ spaces
related by the operation of {\it flopping}\/ \refs{\rTY,\rflops}.
We will review this shortly.
What about the other 95 regions? We will postpone a detailed
answer to this question to section VI and also  to a forthcoming
paper, but there is one essential point worthy of emphasis here.
Some of these regions correspond to \CY\ spaces with orbifold singularities
(similar to those studied in \rDHVW\ except that the covering space
is a nontoroidal \CY\ space). Conventional wisdom and detailed analyses
have always considered such orbifolds to be ``boundary points in
\CY\ moduli space''.  This implies, in particular, that smooth
\CY\ theories are represented by the generic points in moduli space while
orbifold theories are special isolated  points. It also implies that
by giving any nonzero expectation value to ``blow-up modes'' in the orbifold
theory, we move from the orbifold theory to a smooth \CY\ theory.
The present work shows that these interpretations of the orbifold results
are misleading. Rather, it is better to think of orbifold theories
as occupying their own
{\it regions\/} in the enlarged K\"ahler moduli space and hence they are
just as generic as smooth \CY\ theories, which  simply correspond to
other regions.
Furthermore, these orbifold regions are adjacent to smooth \CY\ regions,
but turning on expectation values for
twist fields  does not immediately
resolve the singularities and move one into the \CY\ region. Rather, one must
traverse the orbifold region by turning on an expectation value for
a twist field until one reaches a wall of the smooth \CY\ region. Then,
if one goes further (for which there is no physical obstruction) one
enters the region of smooth \CY\ theories. This is a significant departure from
the hitherto espoused description of orbifolds in string theory.

3) In the mirror manifold construction of \rGP, typically both the
original \CY\
space and its mirror  are singular.  Quite generally, there is more than
one way of repairing these singularities to yield a smooth \CY\ manifold,
and the resulting smooth spaces can be topologically distinct.
A natural question is: is one or some subset of these possible
desingularizations (which are on equal footing mathematically) singled
out by string theory, or are all possible desingularizations realized
by physical models? We show that each possible desingularization to
a smooth \CY\ manifold has its own region in the fully enlarged
K\"ahler moduli space. (In fact, these particular regions constitute
what we call the {\it partially\/} enlarged K\"ahler moduli space.)

\nref\rCGH{P. Candelas, P. Green and T. H\"ubsch, Nucl. Phys. {\bf B330}
(1990) 49.}

4) As remarked earlier, and as will be one of our main foci, the present
work establishes  the veracity of the long suspected belief that string theory
admits physically smooth processes which can result in a change of the
topology of spacetime. Some of the regions in the enlarged K\"ahler moduli
space correspond to the complexified K\"ahler cones of topologically
distinct smooth \CY\ manifolds. Since we show that there is no
physical obstruction to deforming our theories by truly marginal
operators  which take us smoothly from one region to another, we see
that we can change the topology of spacetime in a physically smooth manner.
Furthermore, there is nothing at all exotic about such processes. They
correspond to the most basic kind of deformation  to which one
can subject a conformal field theory. This should be contrasted to
the situation where one can change the topology of a \CY\ manifold by
passing through a {\it conifold\/} point in the moduli space as
studied in \rCGH. In such a process one necessarily encounters singularities.

Our approach to establishing these results \rAGM, as mentioned,
relies heavily on
properties of mirror manifolds originally established in \rGP.
These will be reviewed in the next section. Basically, mirror symmetry
has established that a given conformal field theory may have more than
one geometrical realization as a nonlinear $\sigma$-model with a Calabi-Yau
target space. Two {\it totally different\/} Calabi-Yau spaces can
give rise to isomorphic conformal theories (with the isomorphism
being given by a  change of sign of a
certain charge). One important implication of this
result is that any  physical observable in the underlying conformal theory
has two geometrical interpretations --- one on each of the associated
Calabi-Yau spaces. Furthermore, a one parameter family of conformal field
theories of this sort likewise has two geometrical interpretations in
terms of a family of Calabi-Yau spaces and in terms of a mirror family
of Calabi-Yau spaces. The mirror manifold phenomenon can be an extremely
powerful physical tool because certain questions which are hard to analyze
in one geometrical interpretation are far easier to  address on the mirror.
For instance, as shown in \rGP, certain observables which have
an extremely complicated geometrical realization on one of
the Calabi-Yau spaces (involving an infinite
series of instanton corrections, for example) have an equally simple
geometrical interpretation on the mirror Calabi-Yau space (involving
a single calculable integral over the space).

For the question of topology change,
and more generally, the question of the structure
of \CY\ moduli space, we make use of mirror manifolds in
the following way. The picture we are presenting implies
that under mirror symmetry the complex structure moduli space of $Y$
is mapped to
the {\it enlarged\/} K\"ahler moduli space of $X$ (and vice versa).
{}From this we conclude
that  for any point in the
enlarged K\"ahler moduli space of
$X$ we can find a corresponding point in the complex structure moduli space
of $Y$ such that correlation functions of corresponding observables
are identically equal since these points should correspond to isomorphic
conformal theories.
By choosing representative points which lie in distinct regions of
the enlarged K\"ahler moduli space of
$X$, the veracity of the latter statement provides
 an extremely sensitive test of the picture
we are presenting. In \rAGM\ as amplified upon here, we showed that
this prediction could be explicitly verified in a nontrivial example.

The picture of topology change, therefore, as discussed in
\rAGM\ and here can be summarized as follows.
 We consider a one parameter family of conformal field
theories which have a mirror manifold realization. On one of these two
families of Calabi-Yau spaces, the topological type changes as we progress
through the family since we pass through a wall in the enlarged
K\"ahler moduli space. Classically one would expect to encounter
a singularity in this process. However,
although the classical geometry passes through a singularity, it is
possible that the quantum physics does not\foot{We
emphasize that all of our analysis is
at string tree-level and our use of the term ``quantum'' throughout
this paper refers to quantum properties of the two-dimensional conformal
field theory on the sphere which describes classical string propagation.
It might be more precise to use the word ``stringy'' instead of ``quantum''
but we shall continue to use the latter common parlance.}.
This is a difficult possibility to analyze directly because it is precisely
in this circumstance --- one in which the volume of curves in
the internal space are small --- that we do not trust perturbative methods in
quantum field theory.  However, it
proves extremely worthwhile to consider the description on the mirror
family of Calabi-Yau spaces.
On the mirror family, only the complex structure changes and the K\"ahler form
can be fixed at a large value, thereby allowing perturbation theory
to be reliable.
As we shall discuss,  on the
mirror family
no  topology change occurs --- rather a continuous change of shape
accompanied by  {\it smoothly varying\/} physical observables is all that
transpires. Thus, on the mirror family we can directly see that no
physical singularity is encountered even though one of our geometrical
descriptions involves a discontinuous change in topology\foot{One
might interpret this statement to mean that --- at some level ---
no topology change is really occurring so long as one makes use of the
correct geometrical description. This is not true. A more precise version
of the statement given in the text, and one which will be explained fully
in the sequel, is: certain topology changing transitions associated
with
changing the K\"ahler structure on a Calabi-Yau space can be reformulated
as physically smooth topology preserving deformations of the complex
structure on its mirror.
The general situation is one in which the K\"ahler structure {\it and\/}
the complex
structure of a given Calabi-Yau change and hence similarly for its mirror.
Generically, the K\"ahler structure deformations of both the original
Calabi-Yau and its mirror will result in both families undergoing
topology change.  We can analyze the physical description of such
transitions by studying the mirror equivalent complex structure deformations
(and hence see that the physical description is smooth) but we cannot
give a nonlinear $\sigma$-model geometrical interpretation which does not
involve topology change.}.

To avoid confusion, we emphasize that the topology changing
transitions which   we study are {\it not\/} between a Calabi-Yau
and its mirror. Rather, mirror manifolds are used as a tool to study
topology changing transitions in which, for example, the Hodge numbers
are preserved and only more subtle topological invariants change.
The observation that a given Calabi-Yau space may have a number of
``close relatives'' with the same Hodge numbers but a different
topology (constructed by flopping) was made a number of years ago
by Tian and Yau \rTY.  What is new here is the smooth interpolation
between the $\sigma$-models based on these different spaces.

\nref\rnewW{E. Witten, Nucl. Phys. {\bf B403} (1993) 159.}

Although this picture of topology change, as presented and verified in
\rAGM, is both compelling and convincing, it is natural to wonder how
string theory, at a microscopic level, avoids a physical singularity
when passing through a topology changing transition. The local description
of the topology changing transitions studied here was given in
\rnewW\ which,
contemporaneously with \rAGM, established this first concrete arena
of spacetime topology change.  In \rnewW, by direct examination of
particular correlation functions it was shown that quantum corrections
exactly cancel the discontinuity that is experienced by the classical
contribution --- in precise agreement with what is expected based on \rAGM.
Moreover, the results of \rnewW\ thoroughly and precisely map out the
physical significance of regions in the ``fully enlarged K\"ahler moduli
space'' (which we shall  discuss in some detail).
These  regions are interpreted
in
\rnewW\ as phases of $N = 2$ quantum field theories and shown to include
Calabi-Yau $\sigma$-models on birationally equivalent
but topologically distinct target spaces, \LG\ theories and
other  ``hybrid'' models which we shall discuss in section VI.  The
results of \rnewW\ have helped to shape the interpretations we give here and
provide complimentary evidence in support of the topology changing
processes we present.

\nref\rRoan{S.-S. Roan, Internat. J. Math. {\bf 2} (1991) 439;
``Topological Properties of Calabi-Yau Mirror Manifolds'',
Max-Planck-Institut preprint, 1992.}
\nref\rBatyrev{V. Batyrev, ``Dual Polyhedra and Mirror Symmetry
for Calabi-Yau Hypersurfaces in Toric Varieties'', Essen preprint,
November 18, 1992.}

Much of this paper is aimed at explaining the methods and results of
\rAGM. In section II we will give a more detailed summary of the topology
changing picture established in \rAGM\ while emphasizing the background
material on mirror manifolds that is required. We will see that our
discussion requires some understanding of toric geometry and we will give
a detailed primer on this subject in section III. In section IV we shall
apply some concepts of toric geometry to discuss mirror symmetry following
\rRoan\ and \rBatyrev. We will extend this discussion to yield
a toric description of the K\"ahler and  complex structure moduli spaces ---
naturally leading to the important concepts of the secondary fan ,
the ``partially'' enlarged K\"ahler moduli space and the
monomial-divisor mirror map. In section V we shall apply these concepts
to  verify our picture of moduli space, the action of mirror
symmetry and
 topology change. We will do this in the
context of a particularly tractable example, but it will be clear that our
results are general. We will review the calculation of \rAGM\ which established
the topology changing picture reviewed above. In section VI we will
indicate the structure of the ``fully''
 enlarged moduli space alluded to above and in
\rAGM, and show its relation to the ``hybrid'' models found in \rnewW.
We will leave some important calculations in these theories to a forthcoming
paper. Finally, in section VII we shall offer our conclusions.

\newsec{Mirror Manifolds, Moduli Spaces and Topology Change}

In this section we aim to give an overview of the topology changing picture
established in \rAGM\ and use subsequent sections to fill in essential
technical details that shall arise in our discussion.

\subsec{Mirror Manifolds}
Mirror symmetry was conjectured based upon naturality arguments in
\refs{\rDixon,\rLVW}, was  suggested by the computer studies of
\rCLS\ and was established to exist in certain cases by direction
construction in \rGP. Mirror symmetry describes a situation in which
two very different Calabi-Yau spaces (of the same
complex dimension) $X$ and $Y$, when taken as target spaces for
two-dimensional nonlinear $\sigma$-models, give rise to isomorphic $N = 2$
superconformal field theories (with the explicit isomorphism
involving a change in the sign of a certain $U(1)$ charge).
Such a pair of Calabi-Yau spaces $X$ and $Y$ are said to constitute a
{\it mirror pair\/} \rGP.
Note that the tree level actions of these
$\sigma$-models are thoroughly different as $X$ and $Y$ are topologically
distinct. Nonetheless, when each such action is modified by the
series of corrections required by quantum mechanical conformal invariance,
the two nonlinear $\sigma$-models become isomorphic.

The naturality arguments of \refs{\rDixon,\rLVW}\ were based on the observation
that the two types of moduli in a Calabi-Yau $\sigma$-model --- the K\"ahler
and complex structure moduli
(see the following subsections for a brief review) --- are
very different geometrical objects.
However, their conformal field theory counterparts --- truly marginal
operators --- differ only by the sign of their charge under a $U(1)$ subgroup
of the superconformal algebra. It is unnatural that a pronounced geometric
distinction corresponds to such a minor conformal field theory distinction.
This unnatural circumstance would be resolved if for each such conformal
theory there is a second Calabi-Yau space interpretation in which the
association of conformal fields and geometrical moduli is {\it reversed\/}
(with respect to this $U(1)$ charge)
relative to the first. If this scenario were to be realized, it would imply,
for instance, the existence of pairs of Calabi-Yau spaces whose Hodge
numbers satisfy $h^{p,q}_X = h^{d - p,q}_Y$. A computer survey of hypersurfaces
for the case $d = 3$ \rCLS\ revealed a host of such pairs. It is important
to realize, especially in light of more recent mathematical discussions of
mirror symmetry \refs{\rRoan,\rBatyrev}, that if $X$ and $Y$ satisfy
the appropriate Hodge
number identity  this by no means establishes that they form a mirror pair.
To be a mirror pair, $X$ and $Y$ must correspond to the same conformal
field theory. Such mirror pairs of Calabi-Yau spaces were constructed in
\rGP\ and at present are the only known examples of mirror manifolds. This
construction will play a central role in our analysis so we now briefly
review it.

\nref\rGepner{D. Gepner, Phys. Lett. {\bf 199B} (1987) 380.}
\nref\rGVW{B. Greene, C. Vafa and N. Warner, Nucl. Phys. {\bf B324}
(1989) 371.}
\nref\rMartinec{E. Martinec, in {\it V.G. Knizhnik Memorial Volume}\/
(L. Brink et al., eds.), World Scientific, 1990, p. 389;
Phys. Lett. {\bf 217B} (1989) 431.}

In \rGP\ it was shown that any string vacuum  $K$ built from products of
$N = 2$ minimal models \rGepner\
respects a certain symmetry group $G$ such that
$K$ and $K/{G}$ are isomorphic conformal theories with the explicit
isomorphism being given by a change in sign of the left moving $U(1)$ quantum
numbers of all conformal fields.
Furthermore,
$G$ has a geometrical interpretation  \rGP\ as an action on the
Calabi-Yau
space $X_0$ associated to $K$ \refs{\rGVW,\rMartinec}. This
geometrical action, in contrast to its conformal field theory realization,
does not yield an isomorphic Calabi-Yau space. Rather, $X_0$ and
$Y_0=X_0/{G}$
are topologically quite different. Nonetheless, $K$ and $K/{G}$
are geometrically interpretable in terms of $X_0$ and $Y_0$,
respectively --- and since the former are isomorphic conformal theories,
the latter topologically distinct Calabi-Yau spaces yield isomorphic
nonlinear $\sigma$-models\foot{In light of the results of
\rnewW\ which reveal a subtlety in the Calabi-Yau/\LG\
correspondence, the arguments of \rGP\ more precisely show that
the \LG\  (orbifold) theory corresponding to $K$ and that
associated
to $K/{G}$ are isomorphic.
To arrive at the geometric correspondence between $X_0$ and $Y_0$,
one must vary the parameters in the theory.
This point should become clear in section VI.}.
The two Calabi-Yau spaces therefore
constitute a mirror pair.
As we will discuss in more detail below,
the explicit isomorphism between $K$ and $K/{G}$ being a reversal
of the sign of the left moving $U(1)$ charge implies that
$X_0$ and $Y_0$ have
Hodge numbers  (when singularities are suitably resolved)
which satisfy the mirror relation given in the last paragraph.

\nref\rAorb{P.S. Aspinwall, Commun. Math. Phys. {\bf 128} (1990) 593.}
\nref\rAL{P.S. Aspinwall and C.A. L\"utken,
Nucl. Phys. {\bf B355} (1991) 482.}%

Having built a specific mirror pair of theories, as stressed in \rGP,
one can now use marginal operators to move about the moduli space of
each theory to construct whole families of mirror pairs. It is
important to realize however that this process is being performed at
the level of conformal field theory and that there may be some subtle
difficulties in translating this
into statements about the {\it geometry\/}
of mirror families. In fact, it was shown in
\rAorb\ that there is an apparent contradiction between the
structure of the moduli space of K\"ahler forms according to classical
geometry and according to conformal field theory. This
issue thus carries into mirror symmetry \rAL.
If one builds a mirror pair
of theories by the method of \rGP, then at least one theory will be an
orbifold and the associated target space will have quotient
singularities. Classically, varying parameters in
 the local moduli space of K\"ahler forms
around this point has the effect of ``blowing-up'' these singularities.
Typically this process is not unique and so leads to a fan-like
structure with many regions as will be discussed.
Such a structure is not seen locally in
the mirror partner.
It was suggested in \rAL\ that a resolution of this conundrum would
occur if string theory were somehow able to smooth out so-called
``flops'' which relate different blow-ups to each other.
In this paper we will describe exactly how to to study the geometry of
mirror families. We will see how the fan-like structure appears the
moduli space of conformal field theories once one looks at the global
structure of the moduli space and that flop transitions are indeed
smoothed out in string theory. We will also find that there are
many other transitions that can occur in the fan structure.

\subsec{Conformal Field Theory Moduli Space}

Amongst the operators which belong to a given conformal field theory there
is a special subset, $\{ \Phi_i \}$,
consisting of ``truly marginal operators''.
These operators have
the property that they have conformal dimension $(1,1)$ and hence can be used
to deform the original theory through the addition of terms to the original
action which to first order have the form
\eqn\edeform{ \sum t_i \int d^2\!z\,\Phi_i . }
The $\Phi_i$ being truly marginal, higher order terms can be chosen so that
the resulting theory is still
conformal. We consider all such theories that can be constructed in this manner
to be in the same {\it family\/} --- differing from
each other by truly marginal
perturbations.
The parameter space of all such conformal field
theories is known as the {\it moduli space\/} of the family.

\nref\rYau{S.-T. Yau, Proc. Nat. Acad. Sci. U.S.A. {\bf 74} (1977) 1798.}

If we consider a family of conformal field theories which have a geometrical
interpretation in terms of a family of nonlinear $\sigma$-models, we can
give a geometrical interpretation to the conformal field theory
moduli space. Namely, marginal perturbations in conformal field theory
correspond to deformations of the target space geometry which preserve
conformal invariance. In the case of interest to us, we study $N = 2$
superconformal field theories which correspond to nonlinear $\sigma$-models
with Calabi-Yau target spaces. We will also impose the condition that
the Hodge number $h^{2,0}=0$. There are two types of
geometrical deformations of these
spaces which preserve the Calabi-Yau condition and hence do not spoil
conformal invariance. Namely, one can deform the {\it complex structure\/}
or one can deform the {\it K\"ahler structure}.
In fact, with the above
condition on $h^{2,0}$,
{\it all}\/ of the
truly marginal operators $\Phi_i$ appearing in \edeform\
 have a geometric realization in terms of complex structure
and K\"ahler structure moduli of the associated Calabi-Yau space, and
these two sectors are independent of each other.
The conformal field theory
moduli space is therefore geometrically interpretable in terms of the
moduli spaces parametrizing all possible complex and K\"ahler structure
deformations of the associated Calabi-Yau space, and
is locally a product
of the moduli space of complex structures and the moduli space of
K\"ahler structures.

The truly marginal operators $\Phi_i$ are endowed with an additional quantum
number: their charge under a $U(1)$ subgroup of the $N = 2$ superconformal
algebra. This charge can be $1$ or $-1$ and hence the set of all
$\Phi_i$ can partitioned into two sets according to the sign of this
quantum number. One such set corresponds to the K\"ahler moduli of
$X$ and the other set corresponds to the complex structure moduli of
$X$. It is rather unnatural that such a trivial conformal field theory
distinction --- the sign of a $U(1)$ charge --- has such a pronounced
geometrical interpretation. The mirror manifold scenario removes this
issue in that if $Y$ is the mirror of $X$ then the association of
truly marginal conformal
fields to geometrical moduli is reversed relative to $X$. In this way,
each truly marginal operator has an interpretation as {\it both\/} a
K\"ahler and a complex structure moduli --- albeit on
distinct spaces.

In the next two subsections we shall describe the two geometrical moduli
spaces---the spaces of complex structures and of K\"ahler structures---%
in turn.
Our discussion will, for the most part, be a classical mathematical
exposition of these moduli spaces. It is important to realize that
classical mathematical formulations are generally lowest order
approximations to structures in conformal field theory. The
description of the K\"ahler moduli space given below most certainly
is only a classical approximation to the structure of the corresponding
{\it quantum\/} conformal field theory moduli space.
An important implication of this for our purposes is that if our classical
analysis indicates that a point in the K\"ahler moduli space corresponds
to a singular Calabi-Yau space, it does not necessarily follow that
the associated conformal field theory is singular (i.e.\ has badly behaved
physical observables). Physical properties and geometrical properties
are related but they are not identical.
One might think that a similar statement could be made regarding the
complex structure moduli space --- however  in
our applications there is a crucial difference.
We will only need to study the physical properties of theories represented by
points in the complex structure moduli space which
correspond to smooth (i.e.\ transverse) complex structures.
By Yau's theorem \rYau, in such a circumstance, one can find a smooth
Ricci-flat
metric (which solves the lowest order $\beta$-function equations) which is
in the same cohomology class as any chosen K\"ahler form on the manifold.
By choosing this K\"ahler form to be ``large'' (large overall volume and large
volume for all rational curves) we can trust perturbation theory and
all physical observables are perfectly well defined. Thus,  we can trust
that {\it nonsingular\/} points in the complex structure moduli space
give rise to { \it nonsingular\/}
physics (for sufficiently general choices of the K\"ahler class).
It is this fact which shall play a crucial role in our analysis of
topology changing transitions.

\subsec{Complex Structure Moduli Space}

A given real $2d$ dimensional manifold may admit more than one way
of being viewed as a complex $d$ dimensional manifold. Concretely,
a complex $d$-dimensional manifold is one in
which complex coordinates $z_1,\ldots,z_d$ have been specified in various
``coordinate patches''
such that transition functions between patches are holomorphic functions
of these coordinates. Any two sets of such complex coordinates which
themselves differ by an invertible holomorphic change of variables are
considered equivalent. If there is no such holomorphic change of
variables between two sets of complex coordinates, they are said to
define different complex structures on the underlying real manifold.

\nref\unobstructedness{
F.A. Bogomolov, Dokl. Akad. Nauk
  SSSR {\bf 243} (1978), no.~5, 1101\semi
G. Tian, in {\it Mathematical Aspects of String
Theory}\/ (S.-T. Yau, ed.), World Scientific, 1987, p. 629\semi
A. Todorov, Commun. Math. Phys. {\bf 126} (1989) 325.}%
\nref\rGH{P. Green and T. H\"ubsch, Commun. Math. Phys.
{\bf 113} (1987) 505.}

Given a complex structure on a complex manifold $X$, there is a cohomology
group which parameterizes all possible infinitesimal deformations of
the complex structure: $H^1(X,T)$, where $T$ is the holomorphic tangent
bundle. For the case of $X$ being Calabi-Yau, it has been shown
\unobstructedness\ that
there is no obstruction to integrating infinitesimal deformations to finite
deformations
and hence this cohomology group may be taken as the tangent space to
the parameter
space of all possible complex structures on $X$.
As is well known, because
$X$ has a nowhere vanishing holomorphic $(d,0)$ form, we have the isomorphism
$H^1(X,T) \cong H^{d-1,1}(X)$. For certain types of Calabi-Yau spaces $X$,
there is a simple way of describing these complex structures. In the present
paper we will focus almost exclusively on hypersurfaces in
weighted projective space. So, let $X$ be given as the vanishing locus
of a single homogeneous polynomial in the weighted projective
space $\WCP{(d+1)}{k_0\ldots.,k_{d }}$. Our notation here is that if
$z_0.,\ldots,z_d$ are homogeneous coordinates in this weighted projective space
then we have the identification
\eqn\eid{ [z_0.,\ldots,z_d] \cong [\lambda^{k_0}
z_0,\ldots,\lambda^{k_d}z_d] .}
For $X$ to be Calabi-Yau it must be homogeneous of degree equal to the
sum of the weights $k_i$. Consider the most general form for the defining
equation of $X$
\eqn\eX{ W=\sum a_{i_0 i_1\ldots i_d} z_{i_0}^{p_0}\ldots z_{i_d}^{p_d} =0}
with $\sum k_{i_j}p_j = \sum k_j$.
In order to count each complex structure that arises here only once,
we must make identifications among sets of coefficients $a_{i_0 i_1\ldots i_d}$
which give rise to isomorphic hypersurfaces through
general projective
linear coordinate transformations on the $z_i$.
This gives  a space of possible
complex structures that can be put on the real manifold underlying $X$
(but in general there may be other complex structures as well \rGH).
Thus, we have a very simple description of this part of
the complex structure moduli
space of the Calabi-Yau manifold $X$. We will illustrate these ideas in
an explicit example in section V.

Given a Calabi-Yau hypersurface defined by \eX, note that each monomial
$z_{i_0}^{p_0}\ldots z_{i_d}^{p_d}$ appearing in \eX\ can be regarded
as a truly marginal operator which deforms the complex structure.  These
deformations enter into \eX\ in a purely linear way---no higher
order corrections are necessary.

One subtlety in the above description is that not all choices of the
coefficients $a_{i_0 i_1\ldots i_d}$ lead to a nonsingular $X$.\foot{More
precisely, not all choices lead to an $X$ which is ``no more singular
than it has to be''.  There will be certain singularities of $X$ which
arise from singularities of the weighted projective space itself; we
wish to exclude any {\it additional\/} singularities.} Namely,
if there is a solution to the equations
${\partial f}/{\partial z_i}=0$ other than $z_i=0$ for all $i$,
then $X$ is not smooth. In the moduli space of complex
structures, this singularity condition is met on the ``discriminant locus''
of $X$, which is  a complex codimension one variety. Being complex
codimension one, note
(figure 1) that we can choose a path between
any two nonsingular
complex structures which avoids the discriminant locus. This will be
a useful fact  later on.

\iffigs
\midinsert
$$\vbox{\centerline{\epsfxsize=10cm\epsfbox{catp-f01.ps}}
\centerline{Figure 1. The moduli space of complex structures.}}$$
\endinsert
\fi

\subsec{K\" ahler Structure Moduli Space}

In addition to deformations of the complex structure of $X$, one can
also consider deformations of the ``size'' of $X$.
More precisely, $X$ is a K\"ahler manifold and hence is endowed with
a K\"ahler metric $g_{i \overline\jmath} dz^i \otimes d
{\overline z^{\overline\jmath}}$ from
which we construct the K\"ahler form
$J = ig_{i \overline\jmath} dz^i  \wedge d{\overline
z^{\overline\jmath}}$.
The K\"ahler form
is a closed $(1,1)$ cohomology class, i.e.\ is an element of $H^{1,1}(X)$.
Deformations of the K\"ahler structure on $X$ refer to deformations
of the K\"ahler metric (hence the``size'') which preserve the $(1,1)$
nature of $J$ and which cannot be realized by a change of coordinates
on $X$. (We remark that deformations of the complex structure
{\it do not\/}
preserve the type $(1,1)$ nature of $J$.)
Such deformations yield
new K\"ahler forms $J'$ which are in distinct cohomology classes.
The space of all such distinct $(1,1) $ classes is precisely given by
$H^{1,1}(X)$ whose general member can be written as
\eqn\eKahler{ \sum a_i e_i}
where $a_i$ are real coefficients and $e_i$ are a basis for $H^{1,1}(X)$.

Not every choice of the $a_i$ gives rise to an acceptable K\"ahler form
on $X$. To be a K\"ahler form, the $(1,1)$ form must be such that it gives
rise to positive
volumes for topologically nontrivial curves, surfaces, hypersurfaces, etc.
which reside in $X$. That is, we require
\eqn\epositive{\int_{C_r} J^r > 0}
where $C_r$ is a homologically nontrivial effective algebraic
$r$-cycle and $J^r$ denotes
$J \wedge J \wedge\ldots\wedge J$ (with $r$ factors of $J$). The K\"ahler
moduli space of $X$ is thus given by a cone (the ``K\"ahler cone'') which
consists of those $(1,1)$ forms which satisfy \epositive. This is a real
space whose dimension is $h^{1,1}$.

\nref\rCd{P. Candelas and X.C. de la Ossa, Nucl. Phys. {\bf B342} (1990) 246.}
\nref\rGS{V. Guillemin and S. Sternberg, Invent. Math. {\bf 97} (1989) 485.}
\nref\rFriedman{R. Friedman, Proc. Symp. Pure Math. {\bf 53} (1991) 103.}%

To gain a better understanding of the K\"ahler moduli space, it proves
worthwhile to study \epositive\ in the special case of $r = 1$ --- that is,
the case in which $C_r$ is a curve.
We consider a curve $C$ with the property that the limiting condition
$\int_C J  = 0$ can be achieved while simultaneously maintaining all of
the remaining conditions in
\epositive\ for all\foot{In practice, we may need to allow the condition
$\int_{C_i} J  = 0$ to hold for a finite number of holomorphic curves $C_i$,
all of which lie in the same cohomology class as $C$.
It is crucial for this discussion that there be {\it only}\/ finitely
many such curves.}
 other curves, and for all higher dimensional subspaces.
These inequalities (and equality) define a ``boundary wall'' in the
moduli space. We can approach this
wall by  changing the K\"ahler metric so as to shrink the volume of the curve
$C$ to a value which is  arbitrarily small. The limiting wall is defined as
the place in moduli space where the volume of $C$ has been shrunk to zero ---
one says that $C$ has been ``blown down'' to a  point. What happens if one
goes even further and allows the  $a_i$ in \eKahler\  to take on values which
pass through to the other side of the wall?
Formally, the volume of the curve
$C$ would appear to become negative. This uncomfortable conclusion, though,
can have a very natural resolution. The curve $C$ can actually have positive
volume on
the other side of the wall, however it must be viewed as residing on
a {\it topologically different\/} space\foot{This interpretation of the
volumes is implicit in the analysis of \rCd, which studied the metric behavior
of the conifold transitions.  It has also been considered in the mathematics
literature \rGS.  The {\it topology changing}\/ property of these
transformations was first pointed out by Tian and Yau \rTY; see \rFriedman\
for an update.}.
This procedure of
blowing a curve $C$ down to a point and then restoring it to
positive volume (``blowing up'') in a manner which changes the topology of
the underlying space is known as ``flopping''. We can think of this flopping
operation of algebraic geometry as providing a means of traversing a
wall\foot{It
must be stressed that not every wall of the K\"ahler moduli space
has this property---this only happens for the so-called ``flopping walls''.
There are other walls, at which certain families of
curves also shrink down to zero,
for which the change upon crossing the wall is not of this geometric
type, but rather, a new kind of physical theory is born.  In the simplest
cases, after shrinking down the curves we will have orbifold
singularities, and previous ``blowup modes'' go over into ``twist field
modes''. We will encounter some of these other walls in section VI.} of
the K\"ahler cone by passing through a {\it singular\/} space and then on to
a different smooth topological model.
Typically there may be many algebraic curves on $X$ which define this kind of
wall of the K\"ahler cone and so can be flopped on in this way.
Each of these flopped models has its own K\"ahler cone
with walls determined in the manner just described. Our discussion, therefore,
leads to the natural suggestion that we enlarge our perspective on the
K\"ahler moduli space so as to include {\it all\/} of these K\"ahler
cones glued together along their common walls. We will call this the
``partially enlarged moduli space'' for reasons which shall become clear
in section VI.

We emphasize that in passing through a wall (i.e.\ in flopping a curve)
the Hodge numbers of $X$ remain invariant. Thus, the topology change
involved here is different from that, for example, encountered in the conifold
transitions of \rCGH.
However, more refined topological
invariants do change. For instance, the intersection forms
on  flopped models generally do
differ from one another.
Again, we shall see this explicitly in an example in section V.

So much for the mathematical description of the space of
K\"ahler forms on a Calabi-Yau manifold and on its flopped versions.
Conformal field theory instructs us to modify our picture of the
 K\"ahler moduli space in two important ways.

First, as discussed, there are quantum corrections to this classical
analysis which in general prove difficult to calculate exactly.
We will get on a handle on such corrections by appealing to mirror
symmetry.

Second, we have noted that the  real dimension of the K\"ahler moduli space
is equal to $h^{1,1}$. Conformal field theory instructs us to double this
dimension to {\it complex\/} dimension $h^{1,1}$ by combining our real
K\"ahler form $J$ with the {\it antisymmetric tensor field\/} $B$ to
form a complexified K\"ahler form  $K  = B + iJ$. The motivation for doing this
comes from supersymmetry transformations which show that it is precisely
this combination that forms the scalar component of a spacetime superfield.
Our discussion concerning the conditions on $J$ carries through unaltered ---
being now applied to the imaginary part of $K$. There are no constraints on
the $(1,1)$ class $B$ --- however, the conformal field theory is invariant
under shifts in $B$ by integral $(1,1)$ classes, i.e.\ elements of
$H^2(X, \BZ)$. Incorporating this symmetry naturally leads us to
exponentiate the na\"\i ve coordinates on K\"ahler moduli space and
consider the  true coordinates to be $w_k = e^{2 \pi i (B_k + iJ_k)}$
where $B_k$ and $J_k$ are the components of the two-forms $B$ and $J$
relative to an integral basis of $H^2(X, \BZ)$. In terms of these exponentiated
complex coordinates, the adjacent K\"ahler cones of the partially
enlarged moduli space (figure 2) now become bounded domains
attached along their common closures as illustrated in figure 3.
Note that the walls of the various K\"ahler cones --- and their
exponentiated versions as boundaries of domains --- are real codimension
one and hence divide the partially enlarged moduli space into regions.
It is impossible to pass from one region into another without passing
through a wall.

\iffigs
\midinsert
$$\vbox{\centerline{\epsfxsize=5cm\epsfbox{catp-f02.ps}}
\centerline{Figure 2. Adjoining K\"ahler cones.}}$$
\endinsert
\fi

\subsec{Topology Change}

In this subsection we will consider the implication of applying
the mirror manifold discussion of subsection $2.1$ to the moduli
space discussion of subsections $2.2$--$2.4$.

\iffigs
\midinsert
$$\vbox{\centerline{\epsfxsize=6cm\epsfbox{catp-f03.ps}}
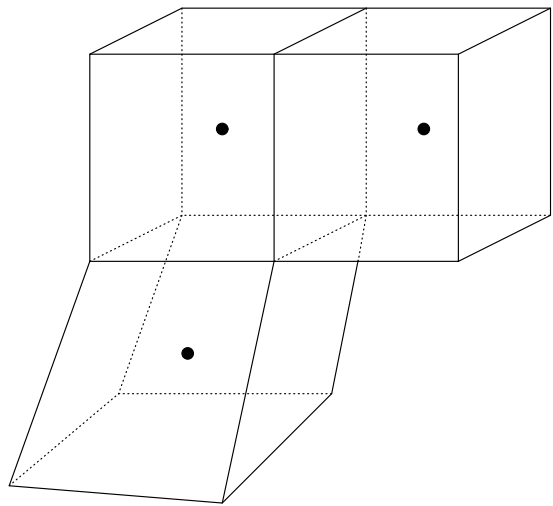
\centerline{Figure 3. Adjoining Cells.}}$$
\endinsert
\fi

As we have discussed, if $X$ and $Y$ are a pair of mirror  Calabi-Yau spaces,
their corresponding conformally invariant nonlinear $\sigma$-models are
isomorphic. We mentioned earlier that the explicit isomorphism involves
a change in sign of the left-moving $U(1)$ charge of the $N = 2$
superconformal algebra. From our discussion of  subsection $2.2$
we therefore see that this isomorphism maps complex structure moduli of
$X$ to K\"ahler moduli of $Y$ and vice versa.  Even as a local result
this is a remarkable statement --- mathematically $X$ and $Y$ are
{\it a priori\/} unrelated Calabi-Yau spaces. Mirror symmetry establishes,
though, a physical link --- their common conformal field theory. Furthermore,
the roles played by the complex and K\"ahler structure moduli  of $X$ and
$Y$ are reversed in the associated physical model. As a global result which
claims an isomorphism between the full quantum mechanical
K\"ahler  moduli space of $X$ and the complex structure moduli space of
$Y$ and vice versa, the statement harbors great potential but also
raises a confusing issue. If we compare figures
1 and 3 we are led to ask:
how is it possible for these two spaces to be isomorphic
when manifestly they have different structures? Specifically,  figure
3 is divided up into cells with the cell walls corresponding to
singular Calabi-Yau spaces lying at the transition between distinct
topological types. In figure 1, however, the space contains no such
cell division. Rather, there are real codimension two subspaces parametrizing
singular complex structures. In figure 3, the passage from one cell to
another necessarily passes through a wall, while in figure 1 --- by
judicious choice of path --- we can pass between any two nonsingular complex
structures without encountering a singularity. So, the puzzle we are faced with
is how are these seemingly distinct spaces isomorphic?

There is another compelling way
of stating this question. As mentioned, in the known construction
of mirror pairs \rGP, typically both of the geometric spaces
$X_0$ and $Y_0$ have quotient
singularities.
We know that string propagation on such singular spaces, say $X_0$, is well
defined \rDHVW\ because the string effectively resolves the singularities.
However, in some situations there is more than one way of repairing
the singularities giving rise to topologically distinct smooth
spaces.
Resolving singularities therefore involves a choice of
desingularization\foot{
Stating this more carefully, bearing in mind that we have actually had
to vary parameters from an initial
Landau-Ginzburg theory to arrive at $X_0$, we are asserting that by
a further variation of parameters $X_0$ can deformed into more
than one \CY\ manifold.}. These manifolds differ by flops \rflops. The moduli
space of \CY\ manifolds
takes the form of figure 3.
(This partially enlarged moduli space does {\it not\/} include the point
corresponding to the Landau-Ginzburg theory but this
will not be important until later in this paper.)
On the mirror
to $X$, namely $Y$, one would expect to find some corresponding choices
in the {\it complex structure\/}
moduli space. In figure 1, however, there are no
divisions into regions, no topological choices to be made. Thus,
what is the physical significance of the topologically distinct
regions of figure 3?

Two possible answers to these questions immediately present themselves, however
neither is at first sight convincing.
 First, it might be that only one
region in figure 3 has a physical interpretation and this region
would correspond under mirror symmetry to the whole complex structure moduli
space of $Y$.
This explanation implies that the operation of flopping rational curves
(passing through a wall in figure 3) has no conformal field theory
(and hence no physical) realization. Furthermore, it helps to
resolve the asymmetry between figures 1 and 3 as neither, effectively,
would be divided into regions.
This explanation would imply that of all the possible
resolutions of singularities of $X_0$ --- which are on completely equal
footing from the mathematical perspective --- the string somehow picks
out one. Although unnatural, {\it a priori\/}
the string might make some physical
distinction between these possibilities.

As a second possible resolution of the puzzle,
it might be that all of the regions
in figure 3 are realized by physical models but, as mentioned, this
is not immediately convincing because the walls dividing the
K\"ahler moduli space in figure 3 into topologically distinct
regions have no counterpart in the complex structure moduli space of the
mirror.
However, it is important to realize that our discussion of figures 1 and
3
has been based on classical mathematical analysis. As discussed earlier,
although we can trust that nonsingular points in the complex structure
moduli space correspond to nonsingular physical models it is generally
{\it incorrect\/} to conclude that singular points in the K\"ahler moduli space
correspond to singular physical models. It is therefore possible that the
quantum version of
figures 2 and 3 {\it are\/} isomorphic with
generic points on the walls of the classical version of figure 3
corresponding to {\it nonsingular\/} physical models. In particular, this
would imply that one can pass from one topological type to another
in figure 3 --- necessarily passing through a singular Calabi-Yau space ---
without encountering a {\it physical\/} singularity.
 Notice that the mirror description of such a process
 does not involve topology change.
Rather, it simply involves a continuous and smooth change in the complex
structure of the mirror space
 (analogous to continuously changing the $\tau$ parameter for
a torus). Thus, this second resolution would establish that
certain topology changing processes (corresponding to flops of
rational curves) are no more exotic than --- and by mirror symmetry can
equally well be described as --- smooth changes in the shape of spacetime.

In \rAGM, we gave compelling evidence that the latter possibility is
in fact correct and we refer to this resolution as giving rise to
{\it multiple mirror manifolds}.
 It is as if a single topological type (focusing on
the complex structure moduli space of $Y$) has not one but many
topological images in its mirror reflection (in the partially enlarged
K\"ahler moduli space of $X$). Mirror manifolds thus yield a rich
catoptric-like moduli space geometry. To avoid confusion, though, note that
a fixed conformal field theory in our family still has precisely two
geometrical interpretations.

We established this picture of topology change in \rAGM\ by verifying
an extremely sensitive prediction of this scenario
which we now review. In section VI
we will also describe a means of identifying the fully enlarged K\"ahler
moduli space of $X$ with the complex structure moduli space of $Y$ by
means of concepts from toric geometry.

\nref\rStromWitten{A. Strominger and E. Witten, Commun. Math. Phys. {\bf 101}
(1985) 341.}
\nref\rStrom{A. Strominger, Phys. Rev. Lett. {\bf 55} (1985) 2547.}%
\nref\rDSWW{M. Dine, N. Seiberg, X.-G. Wen, and E. Witten, Nucl. Phys.
{\bf B278} (1987) 769; Nucl. Phys. {\bf B289} (1987) 319.}%
\nref\rCDGP{P. Candelas, X.C. de la Ossa, P.S. Green, and L. Parkes,
Phys. Lett. {\bf 258B} (1991) 118; Nucl. Phys. {\bf B359} (1991) 21.}%
\nref\rAM{P.S. Aspinwall and D.R. Morrison, Commun. Math. Phys. {\bf 151}
(1993) 245.}%

\def\Y{{Y}}
\def\cM{{X}}
Let $\cM$ and $\Y$ be a mirror pair of Calabi-Yau manifolds. Because
they are a mirror pair, a striking and extremely useful equality between
the Yukawa couplings amongst the $(1,1)$ forms on $M$ and the $(2,1)$ forms
on $\W$ (and vice versa) is satisfied. This equality demands that \rGP:
\eqnn\eEQUAL
$$\displaylines{
\int_{{\Y}} \omega^{abc}
\tilde b^{(i)}_a \wedge \tilde b^{(j)}_b \wedge \tilde b^{(k)}_c
\wedge \omega  = \hfill\eEQUAL \cr
\int_{{\cM}} b^{(i)} \wedge b^{(j)} \wedge b^{(k)} +
\sum_{m,\{u \}}\ex{\int_{\CP1 }u_m^*K}
  \left(\int_{\CP1 } u^*b^{(i)}   \int_{\CP1 } u^*b^{(j)}
 \int_{\CP1 } u^*b^{(k)} \right) .}$$
where on the left hand side (as derived in \rStromWitten)
the $\tilde b^{(i)}_a$ are $(2,1)$ forms (expressed as elements
of $H^1(\Y,T)$ with their subscripts being tangent space indices),
$\omega$ is the holomorphic three form and on the right hand side
(as derived in
\refs{\rStrom,\rDSWW,\rCDGP,\rAM})
the $b^{(i)}$ are $(1,1)$ forms on $\cM$,
$\{u \}$ is the set of holomorphic maps to rational curves on $\cM$,
$u: \CP1 \rightarrow \Gamma $ (with $\Gamma$ such  a holomorphic  curve),
$\pi_m$ is an $m$-fold cover $\CP1 \rightarrow \CP1$ and
$u_m = u \circ \pi_m$.
One should note that the left-hand side of \eEQUAL\ is independent of
K\"ahler form information of $Y$ and the right-hand side is
independent of the complex structure of $X$. In particular \eEQUAL\
can still be valid when $Y$ has rational singularities with K\"ahler
resolutions since such a singular $Y$ may be thought of as a smooth
$Y^\prime$ with a deformation of K\"ahler form.

\nref\rJAMS{D.R. Morrison, J. Amer. Math. Soc. {\bf 6} (1993) 223.}
\nref\otherexamples{
D.R. Morrison, in {\it Essays on Mirror Manifolds}\/ (S.-T. Yau, ed.),
International Press, 1992, p. 241\semi
A. Font, Nucl. Phys. {\bf B391} (1993) 358\semi
A. Klemm and S. Theisen, Nucl. Phys. {\bf B389} (1993) 153\semi
A. Libgober and J. Teitelbaum, Int. Math. Res. Notices (1993) 15\semi
V. Batyrev and D. van Straten, ``Generalized Hypergeometric Functions
and Rational Curves on Calabi-Yau Complete Intersections in Toric
Varieties'', Essen preprint, 1993\semi
P. Candelas, X. de la Ossa, A. Font, S. Katz and D.R. Morrison,
``Mirror Symmetry for Two Parameter Models -- I'', CERN preprint
CERN-TH.6884/93\semi
S. Hosono, A. Klemm, S. Theisen and S.-T. Yau, ``Mirror Symmetry,
Mirror Map and Applications to Calabi-Yau Hypersurfaces'', Harvard preprint
HUTMP-93/0801.}
\nref\rALR{P.S. Aspinwall, C.A. L\"utken, and G.G. Ross, Phys. Lett. {\bf 241B}
(1990) 373.}%

Notice the interesting fact that for $\cM$ at
``large radius'' (i.e., when
$|\exp(\int_{\CP1} u^*_mK)|\ll1$, for all $u$)
the right hand side
of \eEQUAL\ reduces to the topological intersection form on $\cM$ and
hence mirror symmetry (in this particular limit) equates a topological
invariant of $\cM$ to a quasitopological invariant (i.e., one depending on the
complex structure) of $\W$ \rGP.\foot{In
\rCDGP\ equation \eEQUAL\ was combined with an explicit determination of the
{\it mirror map\/} (the map between the K\"ahler moduli space of $\X$ and the
complex structure moduli space of $\X/G$ for $\X$ being the Fermat
quintic in $\CP4$)
to determine the number of rational curves of arbitrary degree on
(deformations of) $\X$. This
calculation has subsequently been described mathematically \rJAMS\
and extended to
a number of other examples \otherexamples.}
In a simple case a suitable limit was found in \rALR\ such that the
intersection form of a manifold could indeed be calculated from its
mirror partner in this way.
If the multiple mirror manifold picture is correct,
and all regions of figure 3 are physically realized,
 the following must hold:
Each of the distinct intersection forms,
which represents the large radius limit of the $(1,1)$ Yukawa couplings
on each of the topologically distinct resolutions of $\X_0$, must be
equal to the $(2,1)$ Yukawa couplings on $\W$ for suitable corresponding
``large complex structure'' limits.
That is, if there are $N$ distinct resolutions of $\X_0$,
one should be able to perform a calculation along the lines
of \rALR\ such that $N$ different sets of intersection numbers are
obtained by taking $N$ different limits.
We schematically illustrate these limits
by means of the marked points in the interiors of the regions
in figure 3.
We emphasize that checking \eEQUAL\ in the large radius limit makes
the calculation much more tractable but it does not in any way compromise
our results. This is an extremely sensitive test of the {\it global\/}
picture of topology change and of moduli space that we are presenting.
Now, actually invoking this equality
requires understanding the precise complex structure limits that correspond,
under mirror symmetry, to the particular large K\"ahler structure limits
being taken.
In the simple case studied in \rALR\ it was possible to use discrete
symmetries to make a well-educated guess at the desired large complex
structure. That example did not admit any flops however. It appears
almost inevitable that any example which has the required complexity
to admit flops will be too difficult to approach along the lines of
\rALR. What we are in need of then is a more sophisticated way of
knowing which large complex limits are to be identified with which
intersection numbers of the mirror manifold.
It turns out that the mathematical machinery of {\it toric geometry\/}
provides the appropriate tools for discussing these moduli spaces and
hence for finding these limit points in the complex structure moduli space.
Thus, in the next section we shall give a brief primer on the subject
of toric geometry, in section IV we shall apply these concepts to
mirror symmetry and in section V we shall examine an explicit example.

\newsec{A Primer on Toric Geometry}

\nref\rOda{T. Oda, {\it Convex Bodies and Algebraic Geometry},
Springer-Verlag, 1988.}
\nref\rFulton{W. Fulton, {\it Introduction to Toric Varieties}, Annals
of Math. Studies, vol. 131, Princeton University Press, 1993.}

In this section we give an elementary discussion of toric geometry
emphasizing those points most relevant to the present work. For more
details and proofs the reader should consult \refs{\rOda,\rFulton}.

\subsec{Intuitive Ideas}

Toric geometry describes the structure of a certain class of geometrical
spaces in terms of simple combinatorial data. When a space admits
a description in terms of toric geometry, many basic and essential
characteristics of the space --- such as its divisor classes, its
intersection form and other aspects of its cohomology --- are neatly coded
and easily deciphered from analysis of corresponding lattices. We will
describe this more formally in the following subsections. Here
we outline the basic ideas.

A toric variety  $V$ over $\IC$ (one can work over other fields but that
shall not concern us here) is a complex geometrical space which
contains the {\it algebraic torus\/} $T = \IC^* \times\ldots\times \IC^*
\cong (\IC^*)^n$ as a dense open subset. Furthermore, there is an action
of $T$ on $V$; that is, a map
 $T \times V \rightarrow V$ which extends the natural
action of $T$ on itself.
The points in $V-T$ can be regarded as limit points for the action of
$T$ on itself; these
serve to give a partial compactification
of $T$. Thus, $V$ can be thought of as a $(\IC^*)^n$ together with
additional limit points which serve to partially (or completely)
compactify the space\foot{As we
shall see in the next subsection, this discussion is a bit
na\"\i ve ---
these spaces need not be smooth, for instance.
Hence it is not enough just to say what points are added ---
we must also specify the local structure near each new point.
}.
Different toric varieties $V$, therefore, are
distinguished by their different compactifying sets. The latter, in turn,
are distinguished
by restricting the limits of the allowed action of $T$ --- and these
restrictions can be encoded
in a convenient combinatorial structure as we now describe.

In the framework of an action $T \times V \rightarrow V$ we can focus
our attention on one-parameter subgroups of the full $T$
action\foot{We
use subgroups depending on one {\it complex\/} parameter.}.
Basically, we follow all possible holomorphic curves in $T$ as they act
on $V$, and ask whether
or not the action has a limit point in $V$.
As the algebraic torus $T$ is a commutative algebraic group,
all of its one-parameter
subgroups  are labeled by points in a lattice $N\cong \BZ^n$ in the following
way. Given $(n_1,\ldots,n_n) \in \BZ^n$ and $\lambda \in \IC^*$ we consider
the one-parameter group $\IC^*$ acting on $V$ by
\eqn\eone{\lambda \times
(z_1,\ldots,z_n) \rightarrow (\lambda^{n_1}z_1,\ldots,\lambda^{n_n}z_n)}
where $(z_1,\ldots,z_n)$ are local holomorphic coordinates on $V$
(which may be thought of as residing in the open dense $(\IC^*)^n$ subset
of $V$). Now, to describe all of $V$ (that is, in addition to $T$)
we consider the action of \eone\ in the limit that
$\lambda$ approaches zero (and thus moves from $\IC^*$ into $\IC$).
It is these limit points which supply the partial compactifications of
$T$ thereby yielding the toric variety $V$.
 The limit points
obtained from the action \eone\ depend upon the explicit vector of
exponents
$(n_1,\ldots,n_n) \in \BZ^n$, but many different exponent-vectors
can give rise to the same limit point.
We obtain different toric varieties by
imposing different restrictions on the allowed choices of $(n_1,\ldots,n_n) $,
and by grouping them together (according to common limit points) in
different ways.

We can describe these
restrictions and groupings in terms of
 a ``fan'' $\Delta$ in $N$ which is a collection
of {\it strongly convex rational polyhedral cones\/} $\sigma_i$ in the real
vector space $N_{\IR}  =N \otimes_{\BZ} \IR$. (In simpler language, each
$\sigma_i$ is a convex
cone with apex at the origin spanned by a finite number
of vectors which live in the lattice $N$ and such that any angle
subtended by these vectors at the apex is $<180^\circ$.)
The fan $\Delta$ is a collection
of such cones
which satisfy the requirement that the face of any cone in $\Delta$ is also
in $\Delta$.
 Now, in constructing $V$, we associate a coordinate patch of
$V$ to each {\it large\/} cone\foot{There
are also coordinate patches for the smaller cones, but
we ignore these for the present.}
 $\sigma_i$ in $\Delta$,
where large refers to a cone spanning an $n$-dimensional
subspace of $N$.
This  patch consists
of $(\IC^*)^n$ together with all the limit points of the action \eone\
for $(n_1,\ldots,n_n) $ restricted to lie in $\sigma_i \cap N$.
There is a single point which serves as the common limit point for
all one-parameter group actions with vector of exponents
$(n_1,\ldots,n_n) $ lying in the {\it interior\/} of $\sigma_i$;
for exponent-vectors on the boundary of $\sigma_i$, additional
families of limit points must be adjoined.

We glue these
patches together in a manner dictated by the precise way in which the
cones $\sigma$ adjoin each other in the collection $\Delta$.
Basically, the patches are glued together in a manner such that the
one-parameter group actions in the two patches agree along the common
faces of the two cones in $\Delta$.
We will
be more precise on this point shortly. The toric variety $V$ is therefore
completely determined by the  combinatorial data of $\Delta$.

\iffigs
\midinsert
$$\vbox{\centerline{\epsfxsize=7cm\epsfbox{catp-f04.ps}}
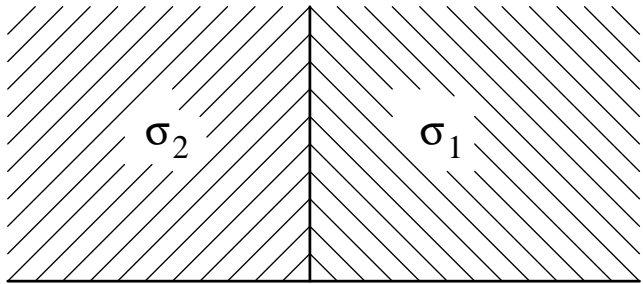
\centerline{Figure 4. A simple fan.}}$$
\endinsert
\fi

For a simple illustration of these ideas, consider the two-dimensional
example of $\Delta$ shown in figure 4. As $\Delta$ consists of two
large cones, $V$ contains two coordinate patches. The first patch ---
corresponding to the cone $\sigma_1$ --- represents $(\IC^*)^2$ together
with all limit points of the action
\eqn\etwo{\lambda \times (z_1,z_2) \rightarrow
(\lambda^{n_1}z_1,\lambda^{n_2}z_2)}
with $n_1$ and $n_2$ non-negative
as $\lambda$ goes to zero.
We add the single point $(0,0)$ as the limit point when $n_1>0$ and
$n_2>0$, we add points of the form $(z_1,0)$ as limit points when
$n_1=0$ and $n_2>0$, and we add points of the form $(0,z_2)$
as limit points when $n_1>0$ and $n_2=0$.\foot{Note
that it is possible to extend this reasoning to the case
$n_1=n_2=0$:  the corresponding trivial group action has as ``limit points''
all points $(z_1,z_2)$, so even the points interior to $T$ itself
can be considered as appropriate limit points for actions by subgroups.}

Clearly, this patch corresponds to $\IC^2$.
By symmetry, we also associate a $\IC^2$ with the cone $\sigma_2$.
Now, these two copies of $\IC^2$ are glued together in a manner
dictated by the way $\sigma_1$ and $\sigma_2$ adjoin each other.
Explaining this requires that we introduce some more formal machinery
to which we now turn.

\subsec{The $M$ and $N$ Lattices}

In the previous subsection we have seen how a fan $\Delta$ in $N_{\IR}$
serves to define a toric variety $V$. The goal of this subsection is
to make the connection between lattice data and $V$ more explicit by showing
how to derive the transition functions between the patches of $V$. The first
part of our presentation will be in the form of an algorithm that
answers: given
the fan $\Delta$ how do we explicitly define coordinate patches and
transition functions for the toric variety $V$? We will then briefly
describe   the mathematics underlying the algorithm.

Towards this end, it proves worthwhile to introduce a second lattice
defined as the dual lattice to $N$, $M=\Hom(N,\BZ)$. We denote the
dual pairing of $M$ and $N$ by $\langle,\rangle$. Corresponding to the fan
$\Delta$ in $N_{\IR}$ we define a collection of dual cones
$\check\sigma_i$
in $M_{\IR}$ via
\eqn\edualc{\check\sigma_i = \{m \in M_{\IR}: \langle
m,n\rangle \ge 0 \hbox{ for all } n
\in \sigma_i \}. }
Now, for each dual cone $\check\sigma_i$ we choose a
finite set of elements $\{m_{i,j} \in M\}$ (with $j=1,\dots,r_i$) such that
\eqn\espan{ \check\sigma_i \cap M = \BZ_{\ge 0 }\, m_{i,1} +\ldots +
 \BZ_{\ge 0}\, m_{i,r_i}.}
We then find a finite set of relations
\eqn\erelations{ \sum_{j = 1}^{r_i} p_{s,j}m_{i,j} = 0}
with $s=1,\dots,R$ such that {\it any\/} relation
\eqn\eanyrelation{ \sum_{j = 1}^{r_i} p_j m_{i,j} = 0}
can be written as a linear combination of the given set, with integer
coefficients.  (That is, $p_j=\sum_{s=1}^R\mu_sp_{s,j}$ for some integers
$\mu_s$.)
We associate a coordinate patch $U_{\sigma_i}$ to the cone
$\sigma_i \in \Delta$ by
\eqn\econstraints{U_{\sigma_i} = \{(u_{i,1},\ldots,u_{i,r_i} )\in \IC^{r_i}
\ |\
u_{i,1}{}^{p_{s,1}}u_{i,2}{}^{p_{s,2}}\ldots u_{i,r_i}{}^{p_{s,r_i}} = 1
\hbox{ for all }s\},}
the equations representing constraints on the variables $u_{i,1}$, \dots,
$u_{i,r_i}$.
We then glue these coordinate patches $U_{\sigma_i}$ and $U_{\sigma_j}$
together by finding a complete set of relations of the form
\eqn\eglue{\sum_{l = 1}^{r_i} q_l m_{i,l}  + \sum_{l = 1}^{r_j} q_l' m_{j,l}
=0}
where the $q_l$ and $q_l'$ are integers.
For each of these relations we impose
the coordinate transition relation
\eqn\etransition{ u_{i,1}{}^{q_1}u_{i,2}{}^{q_2}\ldots u_{i,r_i}{}^{q_r}
u_{j,1}{}^{q_1'}u_{j,2}{}^{q_2'}\ldots u_{j,r_j}{}^{q_r'} = 1.}
This algorithm explicitly shows how the lattice data encodes the defining
data for the toric variety $V$.

Before giving a brief description of the mathematical meaning
behind this algorithm, we pause to illustrate it in two examples.
First, let us return to the fan $\Delta$ given in figure 4.
It is straightforward to see that the dual cones, in this case, take
precisely the same form as in figure 4. We have $m_{1,1} = (1,0),
m_{1,2} = (0,1); m_{2,1} = (-1,0), m_{2,2} = (0,1)$. As the basis
vectors within a given patch are linearly independent, each patch consists
of a $\IC^2$. To glue these two patches together we follow
\eglue\ and write the set of relations
\eqn\eexample{  m_{1,1} + m_{2,1}  =  0}
\eqn\eexamplet{ m_{1,2} - m_{2,2}  =  0.}
These yield the transition functions
\eqn\etransitionex{ u_{1,1}  =  u_{2,1}^{-1} \ , \quad u_{1,2}  =  u_{2,2} .}
These  transition functions imply that the corresponding
toric variety $V$ is the space $\CP1 \times \IC$.

\iffigs
\midinsert
$$\vbox{\centerline{\epsfxsize=5cm\epsfbox{catp-f05.ps}}
\centerline{Figure 5. The fan for $\CP2$.}}$$
\endinsert
\fi

As a second example, consider the fan  $\Delta$ given in figure 5. It is
straightforward to determine that the dual cones in $M_{\IR}$ take the form
shown in figure 6. Following the above procedure we find that the
corresponding toric variety $V$ consists of three patches with
coordinates related by
\eqn\eptwo{ u_{1,1} = u_{2,1}^{-1}\ , \quad u_{1,2} = u_{2,2} u_{2,1}^{-1}}
and
\eqn\eptwot{ u_{2,2} = u_{3,2}^{-1}\ , \quad u_{2,1} = u_{3,1} u_{3,2}^{-1}.}
These transition functions imply that the toric variety $V$ associated
to the fan $\Delta$ in figure 5 is $\CP2$.

\iffigs
\midinsert
$$\vbox{\centerline{\epsfxsize=5cm\epsfbox{catp-f06.ps}}
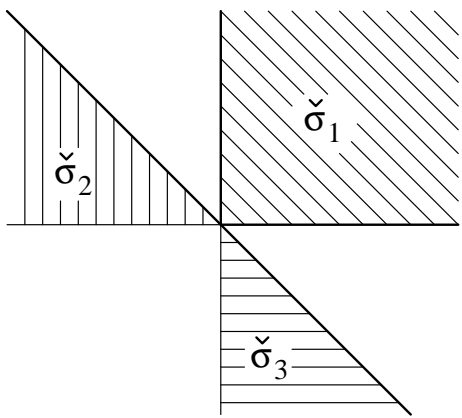
\centerline{Figure 6. The dual cones for $\CP2$.}}$$
\endinsert
\fi

The mathematical machinery behind this association of lattices and
complex analytic spaces relies on a shift in perspective regarding
what one means by a geometrical space. Algebraic geometers
identify geometrical spaces by means of the {\it rings of
functions\/} that are well defined on those spaces.%
\foot{This can be done in a number of different contexts. Different
kinds of rings of functions---such as continuous functions, smooth
functions, or algebraic functions---lead to different kinds of geometry:
topology, differential geometry, and algebraic geometry in the three
cases mentioned.  We will concentrate on rings of algebraic functions,
and algebraic geometry.}
 To make this concrete
we give two illustrative examples. Consider the space $\IC^2$.
It is clear that the ring of functions on $\IC^2$ is isomorphic to
the polynomial ring $\IC[x,y]$ where $x$ and $y$ are formal symbols,
but may be thought of as {\it coordinate functions\/} on $\IC^2$.
By contrast, consider the space $(\IC^*)^2$. Relative to  $\IC^2$,
we want to eliminate geometrical points  either of whose coordinate
vanishes. We can do this by {\it augmenting\/} the ring $\IC[x,y]$ so
as to include functions that are not well defined on such geometrical
points. Namely, $\IC[x,y, x^{-1}, y^{-1}]$ contains functions only
well defined on $(\IC^*)^2$. The ring $\IC[x,y, x^{-1}, y^{-1}]$ can
be written more formally as  $\IC[x,y,z,w]/(zx - 1, wy - 1)$
where the denominator denotes modding out by the  ideal generated by
the listed functions.
One says that
\eqn\especone{ \IC^2 \cong\Spec\IC[x,y] }
and
\eqn\espectwo{ (\IC^*)^2 \cong\Spec{{\IC[x,y,z,w]}\over{(zx - 1, wy - 1)}}}
where the term ``$\Spec$'' may intuitively be thought of as
``the minimal
space of points where the following function ring is well defined''.

With this terminology, the coordinate patch $U_{\sigma_i}$ corresponding
to the cone $\sigma_i$ in a fan $\Delta$ is given by
\eqn\especpatch{ U_{\sigma_i}\cong\Spec\IC[\check\sigma_i \cap M] }
where by $\check\sigma_i \cap M$ we refer to the monomials in
local coordinates that are naturally assigned to lattice points in
$M$ by virtue of its being the dual space to $N$. Explicitly,
a lattice point $(m_1,\ldots,m_n)$ in $M$ corresponds to the monomial
$z_1^{m_1}z_2^{m_2}\ldots z_n^{m_n}$. The latter are sometimes referred
to as {\it group characters\/} of the algebraic group action given by
$T$.

Within a given patch, $\Spec\IC[\check\sigma_i \cap M]$
is a polynomial ring generated by the monomials associated to
the lattice points in $\check\sigma_i$. By the map given in the previous
paragraph between lattice points and monomials, we see that linear
relations between lattice points translate into multiplicative
relations between monomials. These relations are precisely those
given in \erelations.

Between patches,
if $\sigma_i$ and $\sigma_j$ share a face, say $\tau$,
then $\IC[\check\sigma_i \cap M]$ and $\IC[\check\sigma_j \cap M]$
are both subalgebras of $\IC[\check\tau \cap M]$. This provides
a means of identifying elements of $\IC[\check\sigma_i \cap M]$ and
elements of $\IC[\check\sigma_j \cap M]$ which translates into a map
between $\Spec\IC[\check\sigma_i \cap M]$ and
$\Spec\IC[\check\sigma_j \cap M]$. This map is precisely that given in
\etransition.

\subsec{Singularities and their Resolution}

In general, a toric variety $V$ need not be a smooth space. One advantage of
the toric description is that a simple analysis of the lattice data
associated with $V$ allows us to identify singular points. Furthermore,
simple modifications of the lattice data allow us to construct from $V$
a toric variety $\tilde V$ in which all of the singular points are
repaired. We now briefly describe these ideas.

The essential result we need is as follows:

\vskip.2in
\noindent
Let $V$ be a toric variety associated to a fan $\Delta$ in $N$.
$V$ is smooth if for each cone $\sigma$ in the fan we can find a
$\BZ$ basis $\{n_1,\ldots,n_n\}$ of $N$ and an integer $r \le n$ such
that $\sigma = \IR_{\ge 0}\,  n_1 + \ldots  + \IR_{\ge 0}\, n_r$.
\vskip.2in

For a proof of this statement the reader should consult, for
example, \rOda\ or \rFulton.
The basic idea behind the result is as follows.
If $V$ satisfies the criterion in the proposition, then the
dual cone $\check\sigma$ to $\sigma$ can be expressed as
\eqn\edualcone{ \check\sigma = \sum_{i = 1}^r \IR_{\ge 0}\, m_i +
\sum_{i = r+1}^n \IR \, m_i}
where $\{m_1,\ldots,m_n\}$ is the dual basis to $\{n_1,\ldots,n_n\}$.
We can therefore write
\eqn\elattice{\check\sigma \cap M =  \sum_{i = 1}^r \BZ_{\ge 0}\, m_i +
\sum_{i = r+1}^n \BZ_{\ge 0}\, m_i + \sum_{i = r+1}^n \BZ_{\ge 0}\, (-m_i).}
{}From our prescription of subsection $3.2$, this patch is
therefore isomorphic to
\eqn\espece{
\Spec{{\IC[x_1,\ldots,x_n,y_{r+1},\ldots,y_n]}\over
	{\prod_{i=r+1}^n  (x_iy_i - 1 )}}.}
In plain language,
 this is simply $\IC^r \times (\IC^*)^{n-r}$, which
is certainly nonsingular. The key to this patch being smooth is
that $\sigma$ is $r$-dimensional and it is spanned by
$r$ linearly independent lattice vectors in $N$. This implies,
via the above reasoning, that there are no ``extra'' constraints
on the monomials associated with basis vectors in the patch
(see \econstraints)
hence leaving a smooth space.

Of particular interest in this paper will be toric varieties $V$
whose fan $\Delta$ is {\it simplicial}.  This means that each
cone $\sigma$ in the fan can be written in the form
$\sigma = \IR_{\ge 0}\,  n_1 + \ldots  + \IR_{\ge 0}\, n_r$
for some linearly independent vectors $n_1,\dots,n_r\in N$.
(Such a cone is itself called {\it simplicial}.)
When $r=n$, we define a ``volume'' for simplicial cones as follows:
choose each $n_j$ to be the first nonzero lattice point on the
ray $\IR_{\ge0}\, n_j$ and define $\vol(\sigma)$ to be the volume of
the polyhedron with vertices $O,n_1,\dots,n_n$.  (We normalize our
volumes so that the unit simplex in $\IR^n$ (with respect to the
lattice $N$) has volume 1.
Then the volume of $\sigma$ coincides with the index $[N:N_\sigma]$,
where $N_\sigma$ is the lattice generated by $n_1,\dots,n_n$.)
Note that the coordinate chart $U_\sigma$ associated to a simplicial
cone $\sigma$ of dimension $n$ is smooth at the origin precisely when
$\vol(\sigma)=1$.

\iffigs
\midinsert
$$\vbox{\centerline{\epsfxsize=7cm\epsfbox{catp-f07.ps}}
\centerline{Figure 7. The fan for $\IC/\BZ_2$.}}$$
\endinsert
\fi

To illustrate this idea, we consider a fan $\Delta$, figure 7,
which gives rise
to a singular variety. This fan has one big (simplicial)
cone of volume 2 generated by
$v_1 = (0,1) = n_1$ and $v_2 = (2,1) = n_1 + 2n_2$.
The dual cone $\check\sigma$
is generated by $w_1 = (2,-1) = 2m_1 - m_2$ and $w_2 = (0,1) = m_2$.
In these expressions, $n_i$ and $m_j$ are the standard basis vectors.
It is clear that this toric variety is not smooth since it does not
meet the conditions of the proposition. More explicitly, following
\econstraints\  we see that $\check\sigma \cap M = \BZ_{\ge 0}\, (2m_1 - m_2)
+ \BZ_{\ge 0}\, (m_2) +   \BZ_{\ge 0}\, (m_1) $ and hence
$V = \Spec(\IC[x,y,z]/(z^2 - xy = 0) )$. In plain language,
$V$ is the vanishing locus of $z^2 - xy $ in $\IC^3$. This is singular
at the origin, as is easily seen by the transversality test.
 Alternatively, a simple change of variables; $z = u_1u_2$,
$x = u_1^2, y = u_2^2$, reveals that $V$ is in fact $\IC^2/\BZ_2$
(with $\BZ_2$ generated by the action $(u_1,u_2) \rightarrow (-u_1, -u_2)$)
which is singular at the origin as this is a fixed point.
Notice that the key point leading to this singularity is the fact
that we require three lattice vectors to span the two dimensional sublattice
$\check\sigma \cap M$.

The proposition and this discussion suggest a procedure to follow
to modify any such $V$ so as to repair  singularities which it might
have. Namely, we construct a new fan $\tilde \Delta$ from the original
fan $\Delta$ by {\it subdividing}: first subdividing all cones into
simplicial ones, and then subdividing the cones $\sigma_i$ of volume
$>1$ until
the stipulations of the  nonsingularity
proposition are met.
 The new fan $\tilde \Delta$
will then be the toric data for a nonsingular {\it resolution\/} of
the original toric variety $V$. This procedure is called {\it blowing-up}.
We illustrate it with our previous example of $V = \IC^2/\BZ_2$.
Consider constructing $\tilde \Delta$ by subdividing the cone in $\Delta$
into two pieces by a ray passing through the point $(1,1)$. It is then
straightforward to see that each cone in $\tilde \Delta$ meets the
smoothness criterion. By following the procedure of subsection $3.2$
one can derive the transition functions on $\tilde V$ and find that it is
the total space of the line bundle ${\cal O}(-2)$ over $\CP1$
(which is smooth). This is the well known blow-up of the quotient
singularity $\IC^2/\BZ_2$.

If the volume of a cone as defined above
behaved the way one might hope, i.e., whenever dividing
a cone of volume $v$ into other cones, one produced new cones whose
volumes summed to $v$, then subdivision would clearly be a finite
process. Unfortunately this is not the case\foot{If the closest lattice
point to the origin on the subdividing ray does not lie on a face of
the polyhedron $\langle O,n_1,\dots,n_n\rangle$ then the new polyhedra
will be unrelated to the old, and the volumes will not add.}
 and in general one can
continue dividing any cone for as long as one has the patience. This
corresponds to the fact that one can blow-up any point on a manifold to
obtain another manifold. In our case, however, we will utilize the fact that
string theory demands that the
canonical bundle of a target space is trivial. The \CY\ manifold will
not be the toric variety itself as we will see in subsection $3.5$ but we
do require that any resolution of singularities adds nothing new to
the canonical class of $V$.
This will restrict the allowed blowups rather severely.

\nref\rReid{M. Reid, in {\it Journ\'ees de g\'eometrie alg\'ebrique
d'Angers}\/ (A. Beauville, ed.), Sijthoff \& Noordhoff, 1980, p. 273.}

In order to have a resolution which adds nothing new to the canonical
class, the singularities must be what are called {\it canonical
Gorenstein singularities}\/ \rReid.  A characterization of which toric
singularities have this property was given by Danilov and Reid.  To
state it, consider a cone $\sigma$ from our fan $\Delta$, and examine
the one-dimensional edges of $\sigma$.  As we move away
from $O$ along any of these edges we eventually reach a point in $N$. In this
way we
associate a collection of points  $\SS\subset N$ with $\sigma$. (These points
will lie in the boundary of a polyhedron $P^\circ$ which we will
discuss in more detail in subsection $3.5$.)
The fact we
require is that {\it the singularities of the affine toric variety $U_\sigma$
are canonical Gorenstein singularities if
all the points in $\SS$ lie in an affine hyperplane $H$ in $N_{\IR}$ of the
form}
\eqn\hyper{H=\{x\in N_{\IR}\ |\ \langle m,x\rangle=1\} }
{\it for some $m\in M$, and if there are no lattice points $x\in{\sigma\cap N}$
with $0<\langle m,x\rangle<1.$} (\refs{\rReid}, p.294).
Furthermore, in order to avoid adding anything new to the canonical class,
we must choose all one-dimensional cones used in subdividing $\sigma$
from among rays of the form $\IR_{\ge0}\, x$ where $x\in\sigma\cap N$ lies
on the hyperplane $H$ (i.e., $\langle m,x\rangle=1$).

If we have a big simplicial cone, then
the $n$ points in $N_{\IR}$ associated to
the one-dimensional subcones of this cone always define an affine hyperplane in
$N_{\IR}$. If we assume the singularity is canonical and Gorenstein,
then this hyperplane is one integral unit away from the origin and volumes
can be conveniently calculated on it.  In particular,
if the volume of the big cone is greater than
$1$ then this hyperplane
will intersect more points in $\sigma\cap N$. These additional points define
the
one-dimensional cones that can be used for further subdivisions of the
cone that do not affect the canonical class. Since volumes are calculated
in the hyperplane $H$, the volume property
behaves well under such resolutions, i.e., the sum of the volumes of
the new cones is equal to the volume of the original cone that was subdivided.

In some cases, there will not be enough of these additional points to complete
subdivide into cones of volume $1$.  However, in the case
of primary interest in this paper (in which $V$ is a four-dimensional toric
variety
which contains three-dimensional Calabi-Yau varieties as hypersurfaces),
we {\it can\/} achieve a partial resolution of singularities
which leaves only isolated singularities on $V$.  Happily, the Calabi-Yau
hypersurfaces will avoid those isolated singularities, so their singularities
are completely resolved by this process.

\nref{\rmdmm}{P.S. Aspinwall, B.R. Greene and D.R. Morrison,
``The Monomial-Divisor Mirror Map'', IASSNS-HEP-93/43.}

For simplicity of exposition, we shall henceforth assume that our toric
varieties $V$ have the following property: if we partially resolve by means
of a subdivision which (a) makes all cones simplicial, and (b) divides
simplicial cones into cones of volume $1$, adding nothing new to the
canonical class, then we obtain a smooth variety.  This property holds
for the example we will consider in detail in section V. We will point out
from time to time the modifications which must be made when this property
is {\it not}\/ satisfied; a systematic exposition of the general case is
given in \rmdmm.

\nref\rflopsbis{S.-S. Roan,
  Internat. J. Math. {\bf 1} (1990) 211.}

An important point for our study is the fact that, in general, there is
no unique way to  construct
$\tilde \Delta$ from the original
fan $\Delta$.  On the contrary, there are often numerous ways of subdividing
the cones in $\Delta$ so as to conform to the volume $1$ and canonical
class conditions.
Thus, there are numerous smooth varieties that can arise from different ways
of resolving the singularities on the original singular space.
These varieties are birationally equivalent but will, in general, be
{\it topologically distinct}.
For three-dimensional Calabi-Yau varieties such topologically distinct
resolutions can always be related by a sequence of flops \rflops.
For the simplest kind of flops\foot{In fact, these are the only kinds of flops
that we need \rflopsbis.}, a small neighbourhood of the $\CP1$ being flopped
is isomorphic to an open subset of a three-dimensional
toric variety, and that flop can be given a toric description as follows.

\iffigs
\midinsert
$$\vbox{\centerline{\epsfxsize=9cm\epsfbox{catp-f08.ps}}
\centerline{Figure 8. A flop in toric geometry.}}$$
\endinsert
\fi

To a three-dimensional toric variety we associate a fan in $\IR^3$. If
this variety is smooth we can intersect the fan with an $S^2$
enclosing the origin to obtain a triangulation of $S^2$, or part of
$S^2$.
(Different
smooth models will correspond to different triangulations of $S^2$.) We
show a portion of two such triangulations in figure 8. In figure 8 one sees
that if two neighbouring triangles form a convex quadrilateral then
this quadrilateral can be triangulated the other way to give a
different triangulation. Any two triangulations can be related by a
sequence of such transformations. When translated into toric geometry
the reconfiguration of the fan shown in figure 8 is precisely a flop.

\subsec{Compactness and Intersections}

Another feature of the toric variety $V$ which can be directly
determined from the data in $\Delta$ is whether or not it is compact.
Quite simply, $V$ is compact if $\Delta$ covers all of $\IR^n$.
For a more precise statement and proof the reader is referred to
\rOda. This condition on $\Delta$ is intuitively clear. Recall that
we have associated points in $N$ with one parameter group actions on
$V$. Those points in $N$ which also lie in $\Delta$ are special in that
the limit points of the corresponding group actions are part of $V$.
Now, if every point in $N$ lies in $\Delta$ then the limit points
of all one parameter group actions are part of $V$. In other words,
$V$ contains all of its limit points --- it is compact.
The examples we have given illustrate this point. Only the fan of
figure 6 covers all of $N$ and hence only its corresponding
toric variety ($\CP2$) is compact. Note that a
compact toric variety cannot be a \CY\ manifold. This does not stop
toric geometry being useful in the construction of \CY\ manifolds however,
as we shall see.

This picture of complete fans corresponding to compact varieties can
be extended to analyze parts of the fan and gives one a good idea of
how to interpret a fan just by looking at it. If we consider an
$r$-dimensional cone $\sigma$ in the interior of a fan then there is a
$(n-r)$-dimensional complete fan surrounding this cone (in the normal
direction). Thus we can
identify an $(n-r)$-dimensional compact toric subvariety
$V^\sigma\subset V$ associated to
$\sigma$. For example, each one-dimensional cone in $\Delta$ is
associated to a codimension one holomorphically embedded subspace of
$V$, i.e., a {\it divisor}.

We can take this picture further. Suppose an $r$-dimensional cone
$\sigma_r$ is part of an $s$-dimensional cone $\sigma_s$, where $s>r$.
When we interpret these cones as determining subvarieties of $V$ we see that
$V^{\sigma_s}\subset V^{\sigma_r}\subset V$. Now suppose we take two
cones $\sigma_1$ and $\sigma_2$, and find a maximal cone
$\sigma_{1,2}$ such that $\sigma_1$ and $\sigma_2$ are both contained
in $\sigma_{1,2}$. The toric interpretation tells us that
\eqn\eXti{V^{\sigma_{1,2}}\cong V^{\sigma_1}\cap V^{\sigma_2}.}
If no such $\sigma_{1,2}$ exists then $V^{\sigma_1}$ and
$V^{\sigma_2}$ do not intersect. If we take $n$ one-dimensional cones
$\sigma_i$ that form the one-dimensional edges of a big cone then
the divisors $V_{\sigma_i}$ intersect at a point.

Thus we see that the fan $\Delta$ contains information about the
intersection form of the divisors within $V$. Actually the fan
$\Delta$ contains the information to determine self-intersections too
and thus all the intersection numbers are determined by $\Delta$.

Referring back to figure 8 we can describe a flop in the language of
toric geometry. To perform the transformation in figure 8 we first
remove the diagonal bold line the in middle of the diagram. This
line is the base of a two-dimensional cone (which thus has
codimension one).
The only one dimensional compact toric variety is $\IP^1$ so this line
we have removed represented a rational curve. After removing this line
we are left with a square-based cone in the fan which is not simplicial,
so the resulting toric
variety is singular. We then add the diagonal line in the other
direction to resolve this singularity thus adding in a new rational
curve. This is precisely a flop --- we blew down one rational curve
to obtain a singular space and then blew it up with another rational
curve to resolve the singularity.

\subsec{Hypersurfaces in Toric Varieties}

Our interest is not with toric varieties, {\it per se}, but rather
with Calabi-Yau spaces. The preceding discussion is useful in
this domain because a large class of Calabi-Yau spaces can
be realized as hypersurfaces in toric varieties. These were introduced
into the physics literature in \rTY, were argued to
be equivalent, in some sense, to minimal model string vacua in
\refs{\rGepner,\rGVW,\rMartinec,\rnewW}\ and
were  partially cataloged by  computer search (for the case of threefolds)
in \rCLS. The toric varieties of greatest relevance here are
weighted projective spaces. We have seen how ordinary projective
spaces (in particular $\CP2$) are toric varieties and the same is
true for weighted projective spaces.

To illustrate this point, let us construct the weighted projective space
$\WCP{2}{3,2,1}$. As in the case of $\CP2$, there are three patches
for this space. The explicit transition functions between these patches
are:
\eqn\epwtwo{ u_{1,1} = u_{2,1}^{-1}\ , \quad u_{1,2}^3 = u_{2,2} u_{2,1}^{-2}}
and
\eqn\epwtwot{u_{2,2} = u_{3,2}^{-1}\ , \quad u_{2,1}^2 =
u_{3,1}^3u_{3,2}^{-1}.}

\iffigs
\midinsert
$$\vbox{\centerline{\epsfxsize=7cm\epsfbox{catp-f09.ps}}
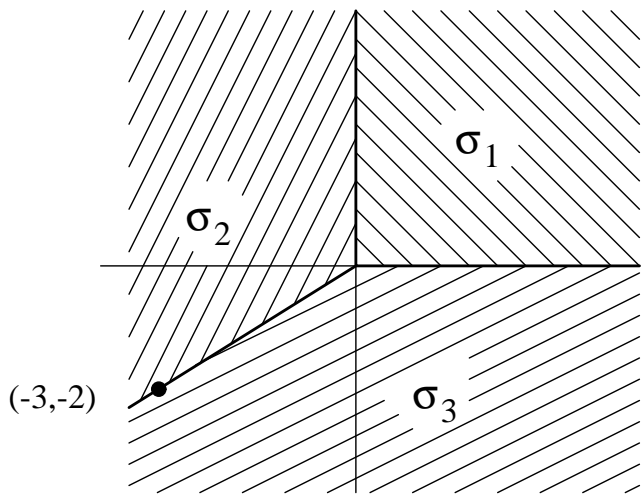
\centerline{Figure 9. The fan for $\WCP{2}{3,2,1}$.}}$$
\endinsert
\fi

Consider the fan  $\Delta$ in figure 9. By following the procedure
of subsection $3.2$ one can directly determine that this fan yields
the same set of transition functions. Notice that
$\WCP{2}{3,2,1}$ is not smooth, by the considerations of subsection $3.2$.
This is as expected since the equivalence relation of
\eid\ has nontrivial fixed points. All higher dimensional weighted
projective spaces can be constructed in the same basic way.

Now, how do we represent a hypersurface in such a toric variety?
In our discussion we shall follow \rBatyrev.
A hypersurface is given by a homogeneous polynomial of degree
$d$ in the homogeneous weighted projective space coordinates.
Recall that points in the $M$ lattice correspond to monomials
in the {\it local\/} coordinates associated to the particular patch
in which the point resides. Consider first the
subspace of $M$ in which all lattice coordinates are positive.
We specify the {\it family\/} of
degree $d$ hypersurfaces by drawing a polyhedron $P$
defined as the minimal convex polyhedron that
surrounds all lattice points corresponding to (the local representation of)
monomials of degree $d$. By sliding this polyhedron along the coordinate
axes of $M$ such that one vertex of $M$ is placed at the origin, we
get the representation of these monomials in the other weighted
projective space patches --- a different patch for each vertex.
To specify a particular hypersurface
(i.e.\ a particular degree $d$ equation) one would need to give more
data than is encoded in this lattice formalism ---
the values of the {\it coefficients\/} of each degree $d$ monomial in
the defining equation of the hypersurface would have to be specified. However,
the toric framework is particularly
well suited to studying the whole family of such hypersurfaces.

As a simple example of this, consider the cubic hypersurface in
$\CP2$ which has homogeneous coordinates $[z_1,z_2,z_3]$. In local
coordinates, say $x = z_1/z_3$ and $y = z_2/z_3$ (the patch in which
$z_3 \ne 0$) the homogeneous cubic monomials are
$1,x,y,x^2,y^2,xy,x^3,y^3,x^2y,xy^2$. (One multiplies each of these
by suitable powers of $z_3$ to make them homogeneous of degree three.)
These monomials all reside in the polyhedral region of $M$ as shown
in figure 10.

\iffigs
\midinsert
$$\vbox{\centerline{\epsfxsize=5cm\epsfbox{catp-f10.ps}}
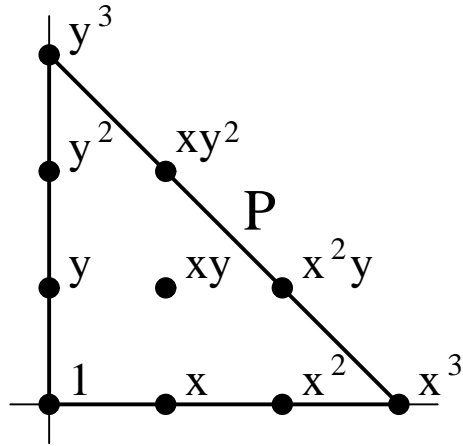
\centerline{Figure 10. The polyhedron of monomials.}}$$
\endinsert
\fi

As shown in \rBatyrev, there is a simple condition on
$P$ to ensure that the resulting hypersurface is Calabi-Yau.
This condition consists of two parts. First, $P$ must contain
precisely one interior point. Second, if we call this interior point
$m_0$ then it must be the case that $P$ is {\it reflexive\/} with
respect to $M$ and $m_0$. This means the following.
Given $P\subset M_{\IR}$ we can construct
the {\it polar polyhedron\/} $P^\circ\subset N_{\IR}$ as
\eqn\eXpp{
P^\circ = \{(x_1,\ldots,x_n)\in N_{\IR};\quad\sum_{i=1}^nx_iy_i
\geq-1\hbox{ for all }(y_1,\ldots,y_n)\in P\},}
where we have shifted the position of $P$  in $M$ so that
$m_0$ has coordinates $(0,\ldots,0)$.
If the vertices of $P^\circ$ lie in $N$ then $P$ is called reflexive. The
origin $O$ of $N$ will then be the unique element of $N$ in
the interior of $P^\circ$.

Also note that given a reflexive $P^\circ \subset N_{\IR}$ we can construct
$\Delta$ by building the fan comprising the cones over the faces,
edges and vertices of $P^\circ$ based on $O$. In general such
a fan may contain cones of volume $>1$. However all the other points
in $P^\circ\cap N$, except $O$, are on faces and edges of $P^\circ$
and can thus be used to resolve the singularities of $V$ without
affecting the canonical class. The exceptional divisors introduced
into $V$ by this resolution of singularities can intersect the
hypersurface to produce exceptional divisors in the \CY\ manifold.
(However, this does not necessarily happen in all cases.
If we consider a point in the
interior of a codimension one face of $P^\circ$ then the exceptional
divisor induced in $V$ would not intersect the hypersurface and
therefore would give
no contribution to the \CY\ manifold.)

To finish specifying $\Delta$ we must say which sets of the one-dimensional
cones are to be used as the set of edges of a larger cone.  Phrased in
terms of the relevant lattice points in $P^\circ\cap N$, what we need
to specify is
{\it a triangulation of $P^\circ$, with vertices in $P^\circ\cap N$,
each simplex of which includes $O$.}  Replacing each simplex with the
corresponding cone whose vertex lies at $O$, we produce the fan $\Delta$.
Conversely, if we are given $\Delta$ then intersecting the cones of $\Delta$
with the polyhedron $P^\circ$ produces a triangulation of $P^\circ$.

\subsec{K\"ahler and Complex Structure Moduli}

Having seen how Calabi-Yau hypersurfaces in  a weighted projective
space are described in the language of toric geometry we now
indicate how the complex structure and K\"ahler structure moduli
on these spaces are represented.

In general, not all such moduli have a representation in toric
geometry. Let's begin with complex structure moduli. As discussed in
subsection $2.3$, such moduli are associated to elements in
$H^{d-1,1}(X)$, where $X$ is the Calabi-Yau space, and under favorable
circumstances \rGH\ some of these can be represented
by monomial perturbations of the same degree of homogeneity as $X$.
By our discussion of the previous subsection, these are the lattice
points contained within $P$. Thus, those complex structure
deformations with a monomial representation have a direct realization
in the toric description of $X$.

The other set of moduli are associated with the K\"ahler structure of
$X$. Note that an arbitrary element of $H^{1,1}(X)$ can, by Poincare
duality, be represented as a $(2d-2)$-cycle in $H_{2d-2}(X)$.
As explained in subsection $3.4$ and above, divisors in $X$ are given by
some of the
one-dimensional cones in $\Delta$ which, in turn, correspond to points
in $P^\circ\cap N$. To be more precise, by this method,
every point in $P^\circ\cap N$
except $O$ and points in codimension one faces of $P^\circ$ gives
a (not necessarily distinct) class in $H_{2d-2}(X)$.

Unfortunately it does not follow that $H_{2d-2}(X)$ is generated by such
points in $P^\circ$. In general an exceptional divisor in $V$ may
intersect $X$ in several isolated regions. This leads to many classes
in $H_{2d-2}(X)$ being identified with the same point in $P^\circ$. If we
define the K\"ahler form on $X$ in terms of the cohomology of $V$ we
thus restrict to only part of the moduli space of K\"ahler forms on
$X$.  We will do this in the next subsection, and study K\"ahler forms
directly on $V$; these always induce K\"ahler forms on $X$\foot{This is
not completely obvious when $V$ is singular, but it is verified in \rmdmm.},
and will produce only part of the K\"ahler moduli space of $X$.
As we will see however, restricting to this part of the moduli space
 will not cause any problems for our
analysis of the mirror property.

\subsec{Holomorphic  Quotients}

\nref\rCox{D.A. Cox, ``The Homogeneous Coordinate Ring of a Toric
Variety'', Amherst preprint, 1992.}

There are two other related ways of building a toric variety $V$
from a fan $\Delta$, in addition to the method we have discussed to
this point.
For a more detailed discussion of
the approach of this subsection the reader is referred to
\rCox.
In this paper we will concern ourselves only with the
holomorphic quotient although another method, the symplectic
quotient, is also quite relevant.

\nref\rAudin{M. Audin, {\it The Topology of Torus Actions on Symplectic
Manifolds}, Progress in Math., vol. 93, Birkh\"auser, 1991.}

An $n$-dimensional toric variety
$V$ can be realized as
\eqn\eintuitholo{( (\IC)^{n+h^{1,1}(V)} - F_{\Delta}) / (\IC^*)^{h^{1,1}(V)}}
where $F_{\Delta}$ is a subspace of $(\IC)^{n + h^{1,1}(V)}$ determined by
$\Delta$. One might wonder why the particular form in
\eintuitholo\ arises. We shall explain this shortly, however, we note that
first, being a toric variety, $V$ contains a $(\IC^*)^n$ as a dense open
set (as in \eintuitholo) and second,
 without removing
$F_{\Delta}$ the quotient is badly behaved (for example it may
not be Hausdorff). For a clear
discussion  of the latter issue we  refer the reader to pages
190--193 of \rnewW.
This is called a holomorphic quotient.
Alternatively, the quotient in \eintuitholo\ can be carried
out in two stages: one can first restrict to one of the level sets of
the ``moment map'' $\mu: (\IC)^{n+h^{1,1}(V)} \to (\IR)^{h^{1,1}(V)}$,
and then take the quotient by the remaining
$(S^1)^{h^{1,1}(V)}$.  (There is a way
to determine which fan $\Delta$ corresponds to each specified value
of the moment map---see for example \rAudin.)
This latter construction is referred to as taking the
symplectic quotient.

The groups $(\IC^*)^k$ by which we take quotients are often constructed
out of a lattice of rank $k$.  If $L$ is such a lattice, we let $L_{\IC}$
be the complex vector space constructed from $L$ by allowing complex
coefficients.  The quotient space $L_{\IC}/L$ is then an algebraic group
isomorphic to $(\IC^*)^k$.  A convenient way to implement the quotient by
$L$ is to exponentiate vectors componentwise (after multiplying by $2\pi i$).
For this reason, we adopt the notation $\exp(2\pi i\,L_{\IC})$ to indicate
this group $L_{\IC}/L$.

Let us consider the holomorphic quotient in greater detail.
To do so we need to introduce a number of definitions.
Let $\AA$ be the set of points in $P^\circ\cap N$.
{\it We assume henceforth that $\AA$ contains no point which lies in the
interior of a codimension one face of $P^\circ$.}
(The more general case is treated in \rmdmm.)
Denote by $r$ the
number of points in $\AA$. Let $\AAO$ be the set $\AA$ with $O$
removed which is
isomorphic to the set of one dimensional cones in the fully resolved
fan $\Delta$.
To every point
$\rho \in \AAO$ associate a formal variable $x_{\rho}$. Let
$\IC^{\AAO} = \Spec\IC[x_{\rho}, \rho \in {\AAO} ]$.
$\IC^{\AAO}$ is simply $\IC^{r-1}$.
Let us define  the  polynomial ideal $B_0$ to be
generated by
 $\{ x^{\sigma}, \sigma\hbox{ a cone in }P^\circ\}$
 with  $x^{\sigma}$ defined as
$\prod_{\rho \notin \sigma} x_{\rho}$.
Let us introduce the lattice $A_{n-1}(V)$ of divisors modulo linear
equivalence on $V$. (On a smooth toric variety, linear
equivalence is the same thing as homological equivalence. See, for
example, \rFulton, p.64, for a fuller explanation.)
This group may also be considered as $H_{2d-2}(V,\BZ)$
if $V$ is compact and smooth which we will assume for the rest of this
section.
Finally, define
\eqn\Gee{G = \Hom(A_{n-1}(V),\IC^*)\cong\exp(2\pi i\,A_{n-1}(V)^\vee)
 \cong (\IC^*)^{h^{1,1}(V)}}
where $A_{n-1}(V)^\vee$ denotes the dual lattice of $A_{n-1}(V)$.
Then, it can be shown that $V$ can be realized as the
holomorphic quotient
\eqn\eholo{ V \cong (\IC^{\AAO}  - F_{\Delta}) / G }
where $F_{\Delta}$ is the vanishing locus of the elements in the ideal
$B_0$.

To give an idea of where this representation of $V$ comes from,
consider the exact sequence
\eqn\eexactoric{ 0 \tto M \tto \BZ^{\AAO} \tto
A_{n-1}(V) \tto 0}
where $\BZ^{\AAO}$ is the free group over $\BZ$ generated by
the points (i.e.\ toric divisors) in $\AAO$. To see why this is
exact, we explicitly consider the maps involved. Elements in
$\BZ^{\AAO}$  may be associated with integer valued
functions defined on the points
in $\AAO$. Any such function, $f$, is given by its value on the
$r-1$ points in $\AAO$. The map from
$\BZ^{\AAO} \rightarrow
A_{n-1}(V)$ consists of
\eqn\emap{f \mapsto \sum_{\rho \in \AAO} f(\rho) D_{\rho} }
where $D_{\rho}$ is the divisor class in $V$ associated to
the point $\rho$ in $\Delta$. Clearly, every toric divisor can be so
written.  It is known \rFulton\ that the toric divisors generate all of
$A_{n-1}(V)$, and hence this map is surjective.

Any element $m \in M$
is taken into  $\BZ^{\AAO}$ by the mapping
\eqn\emapa{m \mapsto \langle \bullet , m\rangle.}
This map
is injective as any two linear functions which agree on
$\AAO$ agree on $N$ (by our assumption that the points of
$\AAO$ span $N$). $M$ is the kernel
of the second map because the points $m \in M$ correspond to global
meromorphic functions (by their group characters, i.e., monomials as
discussed in subsection $3.2$)
and hence give rise to divisors linearly equivalent to 0.

Taking the exponential of the
dual of \eexactoric\ we have
\eqn\eexactorica{1 \tto (\IC^*)^{h^{1,1}(V)}
\tto  (\IC^*)^{\AAO} \tto \exp(2\pi i\,N_{\IC}) \tto
1.}
Notice that  $\exp(2\pi i\,N_{\IC})=N_{\IC}/N$
is the algebraic torus $T$
from which $V$ is obtained by partial compactification. We have now
seen that $T$ arises as a holomorphic quotient of
\eqn\eholob{(\IC^*)^{\AAO}}  by
\eqn\eholoc{ (\IC^*)^{h^{1,1}(V)} .}
To represent $V$ in a
similar manner, therefore, we need to partially compactify this quotient.
The data for so doing, of course, is contained in the fan $\Delta$
(just as it was in  our earlier approach to building $V$). As shown
in \rCox, the precise way in which this partial compactification
is realized in the present setting is to use $\Delta$ to replace
\eholob\ by
the numerator on the right hand side of
\eholo.

\subsec{Toric Geometry of the Partially Enlarged K\"ahler Moduli Space}

The orientation of our discussion of toric geometry to this point has been to
describe the structure of certain Calabi-Yau hypersurfaces.
It turns out that the {\it moduli\/} spaces of these Calabi-Yau
hypersurfaces are themselves realizable as toric varieties.
Hence, we can make use of the machinery we have outlined to not
only describe the target spaces of our nonlinear $\sigma$-model
conformal theories but also their associated moduli spaces.

In this subsection we shall outline how the partially
enlarged K\"ahler moduli space
is realized as a toric variety and in the next subsection
we will do the same for the complex structure moduli space.

First we require a definition for the partially enlarged moduli space.
The region of moduli space we are interested in is the region where one
approaches a large radius limit. Let us therefore partially compactify our
moduli space of complexified K\"ahler forms by adding points
corresponding to large radius limits.

In the discussion above we showed how a toric variety could be
associated to a fan $\Delta$. If the moduli space of the toric variety
is also a toric variety itself then we can describe it in terms of
another fan (in a
different space). This is called the {\it secondary fan\/} and
will be denoted $\Sigma$. $\Sigma$ is a complete fan and thus
describes a compact moduli space.
At first we will not
study the full fan $\Sigma$ but rather the fan $\Psf\subset\Sigma$,
the {\it partial secondary fan}; we describe this fan in detail below. This
fan will specify our partially enlarged moduli space which we will
from now on denote by $\MM_{\Psf}$.

Recall the exact sequence we had for the group $A_{n-1}(V)$ of
divisors on $V$ modulo linear equivalence:
\eqn\eexactmod{0 \tto M \tto \BZ^{\AAO } \tto
A_{n -1}(V) \tto 0.}
By taking the dual and exponentiating we obtained \eexactorica\ and
thus realized $V$ as a holomorphic quotient. Suppose we repeat this
process with \eexactmod\ except this time we {\it do not\/} take the
dual. This will lead to an expression of our toric variety, $\MM_{\Psf}$, as
a compactification of
$(\IC^*)^\AAO/(\IC^*)^n\cong(\IC^*)^{h^{1,1}(V)}$. This is just the
right form for a moduli space of ``complexified K\"ahler forms'' as
discussed earlier in subsection $2.4$.

To actually specify the compactification of the above dense open subset
of $\MM_{\Psf}$, we recall our discussion of subsection $3.2$.
There we indicated that compactifications in toric geometry are
specified by following particular families of one parameter paths
out towards infinity. The limit points of such paths become part of
the compactifying set. The families of paths to be followed are specified
by cones in the associated fan, as we have discussed. This formalism
presents us with a tailor made structure for compactifying the (partially)
enlarged K\"ahler moduli space: take the cones in the associated
fan $\Psf$ to be the K\"ahler cone of $V$ adjoined with the
K\"ahler cones of its neighbours related by flops. The interior of each such
cone, now interpreted as a component of $\Psf$, gives rise to one  point
in the compactification of the partially enlarged K\"ahler moduli space.
This point is clearly the infinite radius limit of the Calabi-Yau space
corresponding to the chosen K\"ahler cone. These are the marked points
in figure 3.

The final point of discussion, therefore, is the construction of the
K\"ahler cone of $V$ and its flopped neighbours. Now, in all
of the applications we shall study, these various birational models
will all arise as different desingularizations of a single underlying
singular variety $V_s$. As discussed in subsection $3.3$,
these desingularizations
can be associated with different fans,
$\Delta$, and are all related by flops of rational curves.
Furthermore, from subsection $3.5$, the construction of $\Delta$ amounts to
a triangulation of $P^\circ$ with vertices lying in the set
$\AAO$. Thus, we expect that the
cones in $\Psf$ will be in some kind of correspondence with the
triangulations based in $\AAO$. We will now describe the precise construction
of
$\Psf$.

To understand the construction of the partial secondary fan, we will need one
technical result
which we now state without proof after
some preliminary definitions.
(The proofs can be found in \refs{\rOda,\rFulton} in the smooth case,
and in \rmdmm\ in the singular case.)
We can consider the intersection of the fan $\Delta$ with $P^\circ$ to
determine a triangulation of $P^\circ$, with vertices taken from
the set of points $P^\circ\cap N$.
We recall that this is a special kind of triangulation (this point
will be important later in this paper). That is, there is a point,
$O$, in the interior of $P$ which is a vertex of every simplex in the
triangulation.
For each $n$-dimensional simplex $\beta$ in $\Delta$ we define
a real linear function by specifying its value at each of the $n$
vertices of $\beta$ except for $O$. The linear function vanishes at $O$.
Let us denote by $\psi_\beta$ such a function defined on
$\beta \in \Delta$. We can extend $\psi_\beta$ by linearity to
a smooth function on all of $N_{\IR}$ which we shall denote by
the same symbol.
Now, we can also define a continuous (but generally not smooth) function
$\psi_{\Delta} : N_{\IR} \rightarrow \IR$ simply by assigning a real number
to each point in $\Delta \cap N$ except $O$ and
within each cone over a simplex, $\beta$, defining
the value of $\psi_{\Delta}$ to be $\psi_\beta$ extended beyond $P$
by linearity. In general this construction
will yield ``corners'' in $\psi_{\Delta}$ at the boundaries between cones.

We say that $\psi_{\Delta}$ is {\it convex\/}
if the following inequality holds:
\eqn\inequal{\psi_\beta(p) \ge \psi_{\Delta}(p)}
for all points $p \in N_{\IR}$.
Similarly, $\psi_\Delta$ is {\it strictly convex\/} if the equality
is true only for points within the cone containing $\beta$.
The theorem alluded to above which we shall need states that
\eqn\eKfms{
\hbox{Space of K\"ahler forms on $V_\Delta$}\cong
{\hbox{Space of strictly convex $\psi_\Delta$}\over\hbox{Space of smooth
$\psi_\Delta$}}.
}
When $V_\Delta$ is singular
we must interpret the ``K\"ahler forms'' in this theorem
in an orbifold sense \rmdmm.

This theorem can also be used to determine whether $V_\Delta$ is K\"ahler
or not. If $V_\Delta$ is not K\"ahler then its K\"ahler cone will be
empty. Thus
\eqn\eXVK{
\hbox{$V_\Delta$ is K\"ahler} \Leftrightarrow \hbox{$\Delta$ admits a
strictly convex $\psi_\Delta$.}}
Such a fan is called {\it regular}.

\nref\rOP{T. Oda and H.S. Park, T\^ohoku Math. J. {\bf 43} (1991) 375.}

Given $\Delta$, then, we can in principle determine the structure of
the K\"ahler cone $\xi_\Delta$ associated to this particular desingularization.
Now consider two smooth toric varieties $X_{\Delta_1}$ and $X_{\Delta_2}$
which are obtained from two different fans $\Delta_1$ and $\Delta_2$
whose intersections with $P^\circ$ give triangulations based on
the same set $\AAO$. This will give two
cones $\xi_{\Delta_1}$ and $\xi_{\Delta_2}$ within the space
$A_{n-1}(V)_\IR$. A function which is strictly convex over $\Delta_1$
cannot be strictly convex over $\Delta_2$ and so
$\xi_{\Delta_1}$ and $\xi_{\Delta_2}$ can only
intersect at their boundaries.
Thus, the K\"ahler cones of different birational models fill out
different regions of $A_{n-1}(V)_\IR$ \rOP.
We can define the partial secondary fan $\Psf$ to consist of all such
cones $\xi_\Delta$, together with all of their faces.

If we take $X_{\Delta_1}$ and $X_{\Delta_2}$ to be related by a flop
then $\xi_{\Delta_1}$ and $\xi_{\Delta_2}$ touch each other
on a codimension 1 wall. One can persuade oneself of this fact by
carefully studying figure 11. The base of the polytope in each case
in this figure
is a section of the fan and the value of $\psi_\Delta$ is mapped out
over this base. The condition that $\psi_\Delta$ is convex is simply
the statement that the resultant surface is convex in the usual sense.
One can move through the space $A_{n-1}(V)_\IR$ by varying the
heights of the solid dots above the base. Note that the flop
transition can be achieved by changing the value of $\psi_\Delta$ at
just one of the points in $\AAO$. (One can mod out by smooth
affine functions by fixing the solid dots at the edge of the base to
be at height zero.)

\iffigs
\midinsert
$$\vbox{\centerline{\epsfxsize=10cm\epsfbox{catp-f11.ps}}
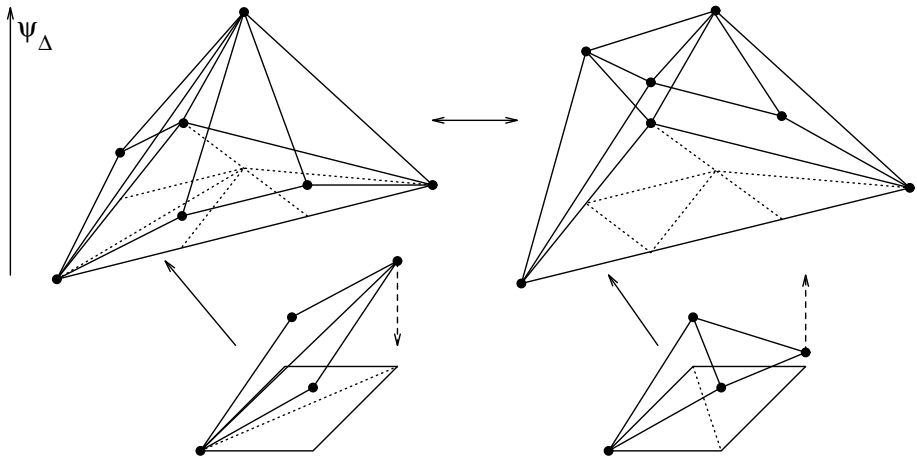
\centerline{Figure 11. A flop in terms of the function $\psi_\Delta$.}}$$
\endinsert
\fi

\nref\rBFS{L.J. Billera, P. Filliman, and B. Sturmfels, Adv. Math.
{\bf 83} (1990) 155.}

This shows that the smooth resolutions of $V_s$
correspond to cones in $A_{n-1}(V)_\IR$ touching each other along
codimension 1 walls if they are related by flops.
We now want to explicitly find these cones.
In practice, the authors of \refs{\rOP,\rBFS} have given a simple algorithm
for carrying out this procedure:

Define an $n\times(r-1)$ matrix $A$ whose columns are the coordinates of
the elements of $\AAO$ in N. Define an integer matrix $B$ as a matrix whose
columns span the kernel of $A$. We will denote the row vectors of $B$
as $b_i$, $i=1\ldots r-1$. These vectors are vectors in the lattice
$A_{n-1}(V)$.

Each big cone $\sigma\in \Delta$ is specified by its one-dimensional subcones
and thus $n$ elements of $\AAO$, say $\rho_i$, $i\in I$. We can then
specify a big cone $\xi_\sigma$ in $A_{n-1}(V)_\IR$ as the cone
which has one-dimensional edges given by $\{b_j\}$ where $j$ runs over
the {\it complement\/} of the set $I$. We then describe the cone
$\xi_\Delta$ associated to $\Delta$ as
\eqn\eXlitc{
  \xi_\Delta = \bigcap\limits_{\sigma\in\Delta}\xi_\sigma.
}

The cones $\xi_\Delta$ for different resolutions of singularities fit
together to form a fan --- the partial secondary fan. As its name
suggests this fan is not complete and thus does not yield a compact
moduli space.
We will explicitly carry out this procedure for a particular example in
section V.

\subsec{Toric Geometry of the Complex Structure Moduli Space}

Let us consider the moduli space of complex structures on a
hypersurface within a weighted projective space.
We will use the $n+1$ homogeneous coordinates $[z_0,\ldots,z_n]$.
If we write down the
most general form of the equation defining the hypersurface (i.e.,
include all terms compatible with the weight of each coordinate) we
obtain something like
\eqn\eGp{
p=a_0z_0z_1\ldots z_n+\ldots+
a_sz_0^{p_0}+a_{s+1}z_1^{p_1}+\ldots+a_dz_d^{n_d}+\ldots+
a_iz_1^{n^i_1}z_2^{n^i_2}+\ldots=0.
}
Let $k$ be the number of terms in this polynomial.
As we vary the complex coefficients $a_i$ we may or may not vary the
complex structure of the hypersurface. Some of the variations in $a_i$
give nothing more than reparametrizations of the hypersurface and so
cannot affect the complex structure. (Moreover, sometimes not all of the
possible deformations of complex structure can be achieved by
deformations of the above type. This always happens for K3 surfaces
for example and can happen in complex dimension 3 if the algebraic
variety has not been embedded
in a large enough ambient space
\rGH.)
A simple reparametrization of the hypersurface is given by the
$(\IC^*)^{n+1}$ action
\eqn\eCdd{
(\IC^*)^{n+1}:(z_0,z_1,\ldots,z_n)\to(\alpha_0z_0,\alpha_1z_1,
\ldots,\alpha_nz_n),\quad\alpha_i\in\IC^*.
}
We will consider the case where the only local deformations\foot{Note
that we have left open
the possibility that there are other, more global deformations which
do not affect the complex structure.  These would take the form of
discrete symmetries preserving the equation \eGp.  We shall ignore
such symmetries for the purposes of this paper; their effects on the
analysis of the moduli space are discussed in detail in \rmdmm.}
 of the
polynomial \eGp\ which fail to give a deformation of complex structure
are deformations which amount to a reparametrization of the form
\eCdd.
One should note that this is quite a strong requirement and
excludes, for example, the quintic hypersurface in $\IP^4$ which has a group of
reparametrizations isomorphic to $Gl(5,\IC)$ rather than $(\IC^*)^5$.
Also, if some of the deformations of complex structure are
obstructed in the sense of \rGH\ (as indeed will happen in our
example) then we only recover a lower dimensional subspace of the
moduli space.

If we first assume that none of the $a_i$'s vanish then we describe
an open subset $\MM_0$ of our moduli
space of the form $(\IC^*)^k/(\IC^*)^{n+1}\cong (\IC^*)^{k-n-1}$. By allowing
some of the $a_i$'s to vanish we can (partially) compactify this space. It
would
thus appear that our moduli space is a toric variety.

It turns out
that to exhibit mirror symmetry most straight-forwardly we need
to modify the above analysis slightly. Let us impose the condition
\eqn\eaisone{a_0=1.}
This reduces the $(\IC^*)^{n+1}$ invariance to $(\IC^*)^n$ --- we have
used the other $\IC^*$ to rescale the entire equation, in setting
$a_0=1$.  Now as
explained earlier, each monomial in \eCdd\ is represented by a point
in $P$ in the lattice $M$. The condition that all reparametrizations are
given by \eCdd\ may be stated in the form that there are no points
from $P\cap M$ in the interior of codimension one faces on $P$.
It can be
seen that the $(\IC^*)^n$ action on any monomial is given by the
coordinates of this point in $M$. This gives
rise to the following exact sequence
\eqn\ecses{
1\tto \exp(2\pi i\,N_{\IC})\tto(\IC^*)^{k-1}\tto\MM_0\tto1
}
which gives another description of $\MM_0$.
The $(\IC^*)^{k-1}$ is the space of polynomials with non-zero coefficients
and $\MM_0$ is the resultant open subset of the moduli space in which no
coefficient vanishes. This
open set can then be compactified by adding suitable regions derived
from places where some of the coefficients vanish. Thus we have again
arrived at something resembling a toric variety.

\newsec{Mirror Manifolds and Toric Geometry}

In the previous section we have given some background on how one
realizes certain families of Calabi-Yau spaces in the formalism
of toric geometry. As is clear from that discussion, many of the detailed
properties and desired manipulations of these spaces are conveniently
encoded in combinatorial lattice data. We now describe how aspects
of mirror symmetry can also be formulated using toric methods.

\subsec{Toric Approach to Mirror Manifolds}

It was originally discovered by S-S. Roan \rRoan\ that the mirror manifold
construction of \rGP\ has a simple and natural description in
toric geometry. Roan found that when the orbifolding occurring in \rGP\
was described in toric terms, it led to an identification between
the $N$ lattice of $X$ and the $M$ lattice of its mirror $Y$.
{}From this,
he could show mathematically that the Hodge numbers of the pairs constructed
in \rGP\ satisfy the appropriate equalities. The results of
Roan, therefore,
 indicate that toric methods provide the correct mathematical language
to discuss mirror symmetry.

After Roan's work, Batyrev \rBatyrev\
has further pursued the application of toric
methods to mirror symmetry and successfully generalized Roan's results.
Batyrev's idea is based on the fact, discussed in
section III, that for a Calabi-Yau
hypersurface in a toric variety the polyhedron
$P$ in the $M$ lattice contains
the data associated with the complex structure deformations and the
polar polyhedron $P^\circ$ in the $N$ lattice contains data associated with
the K\"ahler structure. Since mirror symmetry interchanges these data
it is natural to suspect that if $X$ and $Y$ are a mirror pair,
and if each has a realization as  a toric hypersurface, then
the polyhedron $P$ associated to $X$, say $P_X$,
and its polar $P^\circ_X$ should be isomorphic to $P^\circ_Y$
and $P_Y$, respectively.
In fact,  Batyrev   has shown that for any Calabi-Yau hypersurface  $X$ in
a toric variety described by the  polyhedra $P$ and $P^\circ$ in
$M$ and $N$ respectively, if we construct a new hypersurface $Y$
by interchanging the roles of $P$ and $P^\circ$ then the result is
also Calabi-Yau and furthermore has Hodge numbers consistent with
$Y$ being the mirror of $X$.
This result of Batyrev agrees with that of Roan in the special case
of quotients of Fermat-type hypersurfaces
$X$ and $Y$ being related by orbifolding, but goes well beyond this
class of examples.  It must be borne in mind, though, that true mirror symmetry
involves much more than these equalities between Hodge numbers.
While it seems quite possible that the new pairs constructed by Batyrev
are mirrors, establishing this would require showing that both
members of a proposed pair correspond to isomorphic conformal theories.
This, as yet, has not been done.
We therefore confine our attention to the use of toric methods for
those examples in which the latter conformal field theory
requirement has been established --- namely those of \rGP.

\subsec{Complex Structure vs.~K\"ahler Moduli Space: A Puzzle}

In subsections $3.8$ and $3.9$ we described the K\"ahler and complex
structure moduli spaces of a Calabi-Yau hypersurface in a toric variety
using toric geometry, as these moduli spaces themselves are toric.
If $X$ and $Y$ are a mirror pair, then the K\"ahler moduli space of $X$
must be isomorphic to the complex structure moduli space of $Y$ and
vice versa. This statement, however, immediately presents
us with the puzzle discussed in subsection $2.5$. Namely, the K\"ahler
moduli space is partitioned into cells with the walls of the partitions
corresponding to singular geometrical configurations in which a curve
has been blown down to a point.
 From the point of view of toric geometry,
these cells correspond to the different cones
in the partial secondary fan of subsection $3.8$. The locus of points in the
complex structure moduli space which correspond to singular geometrical
configurations, however, has a very different character. Rather than
being real codimension one, and hence partitioning the moduli space, they
are complex codimension one and hence can generically be avoided when
traversing a path between any two points in the moduli space. How, therefore,
can the two spaces of figures 1 and 3 be isomorphic?

There are two possible answers to this question:

1) It might be that only one region of figure 1 has a physical
interpretation and, under mirror symmetry, corresponds to the whole complex
structure moduli space. This  helps to
resolve the puzzle as we can traverse a
nonsingular path between any two points in a given K\"ahler region just
as we could in the complex structure moduli space. Such a resolution of this
puzzle
would imply that flops have no physical interpretation, that
the string  somehow picks out one of the many possible
resolutions of singularities and that, unfortunately, topology change
in string theory cannot be realized in this manner.

2) The
second possible resolution is that {\it all\/} of the regions in figure 1
have a physical realization and that a generic point on a cell wall
(which corresponds under mirror symmetry to a perfectly smooth complex
structure on $Y$) although geometrically singular, is physically smooth.
This resolution of the puzzle would therefore imply that flops have
a physical realization and, in fact, would provide us with an operation
for changing the topology of spacetime in a physically smooth manner.

The purpose of \rAGM\ and the present work is to present strong evidence
that resolution number two is correct. Let us, therefore, elaborate on this
possibility.

\nref\rW{E. Witten, Nucl. Phys. {\bf B268} (1985) 79;
in {\it Essays on Mirror Manifolds}, (S.-T. Yau, ed.),
International Press Co., 1992, p. 120.}%
\nref\rDG{J. Distler and B. Greene,  Nucl. Phys.  {\bf B309} (1988) 295.}%

 We recall from our discussion in section II that the
mathematical descriptions we
have given for the  complex and K\"ahler moduli spaces in general
are lowest order approximations to the true conformal field theory
moduli spaces.
Therefore, a more precise statement of the isomorphism of moduli spaces
following from mirror symmetry is: the sector of the conformal field
theory moduli space whose lowest order description is
given by  the K\"ahler moduli space
of $X$ is isomorphic to the sector of the conformal field theory
moduli space whose lowest order description  is
given by the complex structure of $Y$, and vice
versa. Now, due to certain nonrenormalization theorems \refs{\rW,\rDG},
the complex
structure moduli space actually describes the corresponding sector of
conformal field theory moduli space exactly. Even without such a result,
we can always focus on a regime in which the volume of $Y$ is large
(corresponding to a point of large complex structure on $X$) and hence
be able to trust quantum field theory perturbation theory. The latter assures
us that nonsingular choices for the complex structure of $Y$ yield
physically well behaved conformal field theories. On the other hand, near
a wall within the K\"ahler moduli space of $X$ perturbation theory
becomes an increasingly poor guide to the physics as the volume of
certain rational curves becomes small. Thus, we cannot trust the classical
result that the manifold becomes singular at the wall to necessarily
indicate that the physics becomes singular as well.
Resolution number two exploits this possibility to the fullest.

How can we choose between these two possible solutions to the issue we
have raised? There is a very delicate prediction which emerges from
solution (2) but not from (1):  solution (2) implies that each point in
figure 3 corresponds, under mirror symmetry, to some point in the
complex structure moduli space of $Y$ (we imagine keeping the complex
structure of $X$ and the K\"ahler structure of $Y$ fixed).
Correlation functions
calculated in the two corresponding nonlinear $\sigma$-models should therefore
be identically equal (if the corresponding operators have been normalized
identically). Thus, a sensitive test of solution (2) is to calculate
a set of correlation functions  at a particular point
in each of the cells of the
K\"ahler moduli space and show that there are corresponding points in the
complex structure moduli space of the mirror such that correlation functions
of corresponding observables (under the identifications following from
mirror symmetry) give identical answers. A specific example of an equality
which follows from this reasoning is given in eqn \eEQUAL.

Ideally, to
carry out this important test we  would understand
the precise map between points in the K\"ahler moduli space of $X$ to
points in the complex structure moduli space of $Y$ (the ``mirror map'')
and also the precise map between observables in the theory based on $X$ to
those of the theory based on $Y$. In reality, though, we can carry out this
test without such precise knowledge of the mirror map, so long as we
carefully choose the points in the K\"ahler moduli space around which we
do our calculation. Namely, if we choose the large radius points in each
of the K\"ahler cells (corresponding to large radius Calabi-Yau
manifolds which are birationally equivalent but topologically distinct)
then toric geometry, as we shall now discuss, provides us with a tool
for locating the corresponding ``large complex structure'' points on the mirror
and also for mapping between observables in the two models.
We emphasize that solution (2), in contrast with (1), implies that the
equalities of correlation functions we have mentioned follows for
{\it any and all\/} choices of points in the partially enlarged K\"ahler
moduli space (that is, including all of the bordering regions we have
been discussing). Our choice to work near large radius points is therefore
one of convenience (calculations are easier there) and in no way
compromises or yields an approximate result.

\subsec{Asymptotic Mirror Symmetry and The Monomial-Divisor Mirror Map}

Given a specific manifold $X$ at some large radius limit our aim is to
determine precisely which ``large complex structure limit'' its mirror
partner $Y$ has attained. This can be achieved if we can find the
mirror map between the complexified K\"ahler moduli space of $X$ and the moduli
space of complex structures of $Y$.
We have already seen in the preceding sections that
in the cases we are considering, both these spaces are isomorphic to
toric varieties which are (compactifications of)
$(\IC^*)^\AAO/(\IC^*)^n$.

In the case of the moduli space of complexified K\"ahler forms on $X$,
$\AAO$ represented the set of toric divisors on $X$. The $(\IC^*)^n$
action represents linear equivalence and is determined by the
arrangement of the points corresponding to $\AAO$ in the lattice $N$.
In the case of the moduli space of complex structures on $Y$, using
the results of subsections $3.9$ and $4.1$, $\AAO$ now represents the set of
monomials in the defining equation for $Y$ (with the exception of the
$a_0$ term) and the $(\IC^*)^n$ action represents reparametrizations
determined by the arrangement of the points of $\AAO$ in the lattice
$N$.

\nref\rpapIII{P.S. Aspinwall, B.R. Greene and D.R. Morrison,
``Limit Points in the Moduli Space of N=2 Conformal Sigma Models'',
IASSNS-HEP-93/49, in preparation.}

We have thus arrived at a natural proposal for the mirror map,
namely to simply identify the divisors of $X$ given by $\AAO\subset N$
with the monomials of $Y$ also given by $\AAO\subset N$. The induced map
between the moduli spaces is called
the {\it monomial-divisor map}, and it is unique up to symmetries of
the point set $\AAO\subset N$.
However, it turns out that although
this proposal for a
mirror map has the correct asymptotic behavior near the large radius
limit points, it unfortunately
differs from the actual mirror map
away from large radius limits. More will be explained concerning
this point in \rpapIII. This problem was also observed from a
different view-point in \rnewW\ where extra small-scale instantons
appeared in the ``A-model''. Fortunately our only concern in this
paper is with large radius limits and so we may take this na\"{\i}ve
identification of the two moduli spaces as an approximation of
 the true mirror map which is adequate for our purposes.
For a more mathematical discussion of these points see \rmdmm.

In order to determine the large radius limits of $X$ we now
consider compactifications of $(\IC^*)^\AAO/(\IC^*)^n$. In terms of
the K\"ahler form, $J$, on $X$ we are studying a limit in which $e^{2\pi
i(B+iJ)}\to0$ by taking $J\to\infty$ inside the K\"ahler cone, $\xi_X$,
of $X$.\foot{More precisely, $\xi_X$ represents that part of the K\"ahler
cone of $X$ which comes from the ambient space $V$ --- it is in reality
the K\"ahler cone of $V$ that we study.}
 In the language of toric geometry, this point added to the
moduli space is given by the cone $\xi_X\subset A_{n-1}(V)_\IR$. In
this way we determine a compactification of the
space of complexified K\"ahler forms on $X$ which includes all large
radius limits. It is the toric variety given by the K\"ahler cone of
$X$ and its neighbours with respect to the lattice $A_{n-1}(V)$.

A large radius limit of $X$ can now be translated into a large complex
structure limit of $Y$. The fact that $J$ remains within $\xi_X$
dictates the relative growth of the coefficients $a_i$ of the
monomials as they are taken to $\infty$ (or 0).
Any path in the moduli space with the property
 that the coefficients of the corresponding
family of hypersurfaces obey these growth properties will approach the
large complex structure limit point specified by $\xi_X$.
We will now demonstrate how
this can be done explicitly by an example.

\newsec{An Example}

In the previous section we described the application of toric geometry
to mirror symmetry and showed how these methods could be used to
verify or disprove our claim that the mathematical operation of flopping
rational curves --- which can result in the change of topology of the
underlying space --- is physically realized and perfectly well behaved.
We will now carry out this procedure in a specific example. We emphasize that
although we work in a specific example of a particular Calabi-Yau and its
mirror (chosen for their relative calculational simplicity) our results
are certainly general. Namely, the operation of flopping is a local
operation which is insensitive to global properties of the space. If we can
show that flopping is physically sensible and realizable in a specific
example, it will certainly have these desired properties in any other
example as the local description and hence physical phenomenon will be
the same.

\subsec{A Mirror Pair of Calabi-Yau Spaces}

We require an example sufficiently complicated to exhibit flops. We
also imposed a condition in subsection $3.9$ to obtain a toric structure for
the
moduli space. That is, we require the group of reparametrizations of
the hypersurface equation to be $(\IC^*)^5$. The following example
is one of the simplest cases meeting these requirements.

Let $X_s$ be a hypersurface in $V\cong\WCP{4}{6,6,3,2,1}$ given by
\eqn\eXfm{f=z_0^3+z_1^3+z_2^6+z_3^9+z_4^{18}=0.}
This space has quotient singularities inherited from
$\WCP{4}{6,6,3,2,1}$. Namely, there are two curves of $\BZ_2$ and
$\BZ_3$ singularities respectively and these intersect at three points
which locally have the form of $\BZ_6$ singularities.
These singularities are the same as the singularities studied in \rAorb.
Any blow-up of
these singularities to give a smooth $X$
gives an exceptional divisor with 6 irreducible
components, thus $h^{1,1}(X)=7$. When one resolves the singularities in
$\WCP{4}{6,6,3,2,1}$ one only obtains an exceptional set with 4
components. One of these components intersects $X$ in regions around
the 3 former $\BZ_6$ quotient singularities. Thus 3 elements of
$H^2(X)$ are being produced by a single element of $H^2(V)$.

In terms of K\"ahler form moduli space one can picture this as
follows. Each of the three $\BZ_6$ quotient singularities contributes a
component of the exceptional divisor. As far as the K\"ahler cone of $X$
is concerned the volume of these three divisors can be varied
independently. If we wish to describe the K\"ahler form on $X$ in
terms of a K\"ahler form on $V$ however, these three volumes had
better be the same since they all come from one class in $H^2(V)$.
Thus we are restricting to the part of the moduli space of K\"ahler
forms on $X$ where these three volumes are equal. An important point
to notice is that even though we are ignoring some directions in
moduli space, we can still get to a large radius limit where {\it all\/}
components of the exceptional divisor in $X$ are large.

The toric variety $\WCP{4}{6,6,3,2,1}$ is given by complete fan around
$O$ whose one dimensional cones pass through the points
\eqn\eXvx{\eqalign{
\alpha_5&=(1,0,0,0)\cr
\alpha_6&=(0,1,0,0)\cr
\alpha_7&=(0,0,1,0)\cr
\alpha_8&=(0,0,0,1)\cr
\alpha_9&=(-6,-6,-3,-2).\cr}}
This data uniquely specifies the fan in this case. (The reason for the
curious numbering scheme will become apparent.) This fan is comprised of
five big cones most of which have volume $>1$. For example, the cone
subtended by $\{\alpha_5,\alpha_7,\alpha_8,\alpha_9\}$ has volume 6.
The sum of the volumes of these 5 cones is 18 and thus we need to
subdivide these 5 cones into 18 cones to obtain a smooth \CY\
hypersurface. The extra points on the boundary of $P^\circ$
which are available to help us do this are
\eqn\eXint{\eqalign{
\alpha_1&=(-3,-3,-1,-1)\cr
\alpha_2&=(-2,-2,-1,0)\cr
\alpha_3&=(-4,-4,-2,-1)\cr
\alpha_4&=(-1,-1,0,0).\cr}}
Note that, as required, none
of these points lies in the interior of a codimension
one face of $P^\circ$.
Any complete fan $\Delta$ of simplicial cones having {\it all\/}
of the lines through
$\{\alpha_1,\ldots,\alpha_9\}$ as its set of one-dimensional cones will
consist of 18 big cones and specify a smooth \CY\ hypersurface, but the
data $\{\alpha_1,\ldots,\alpha_9\}$ does not uniquely specify this
fan.

A little work shows that there are 5 possible fans consistent with
this data, all of which are regular. That is, all 5 possible toric
resolutions of $\WCP{4}{6,6,3,2,1}$ are K\"ahler. We can uniquely
specify the fan $\Delta$ just by specifying the resulting
triangulation of the face $\{\alpha_7,\alpha_8,\alpha_9\}$. The
possibilities are shown in figure 12 and in figure 13 the
three-dimensional simplices are shown for resolution $\Delta_1$.

\iffigs
\midinsert
$$\vbox{\centerline{\epsfxsize=10cm\epsfbox{catp-f12.ps}}
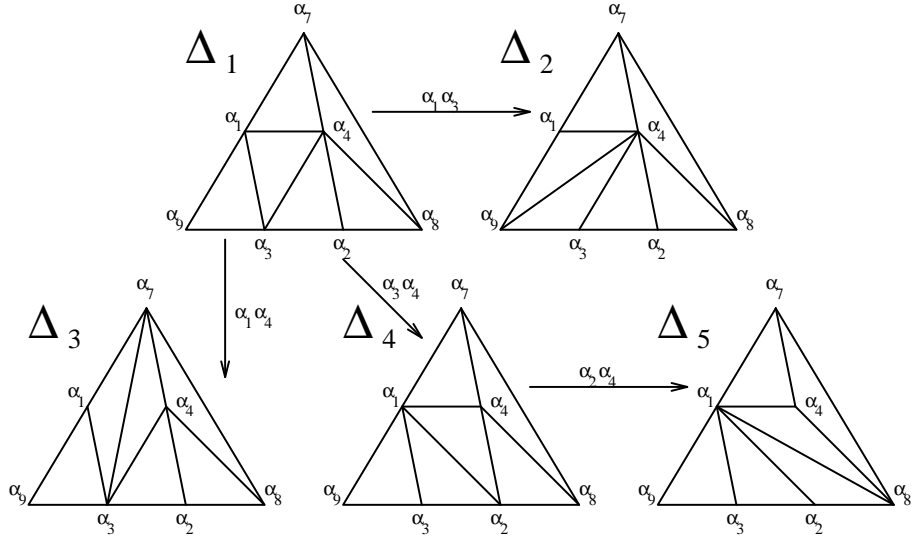
\centerline{Figure 12. The five smooth models.}}$$
\endinsert
\fi

\iffigs
\midinsert
$$\vbox{\centerline{\epsfxsize=10cm\epsfbox{catp-f13.ps}}
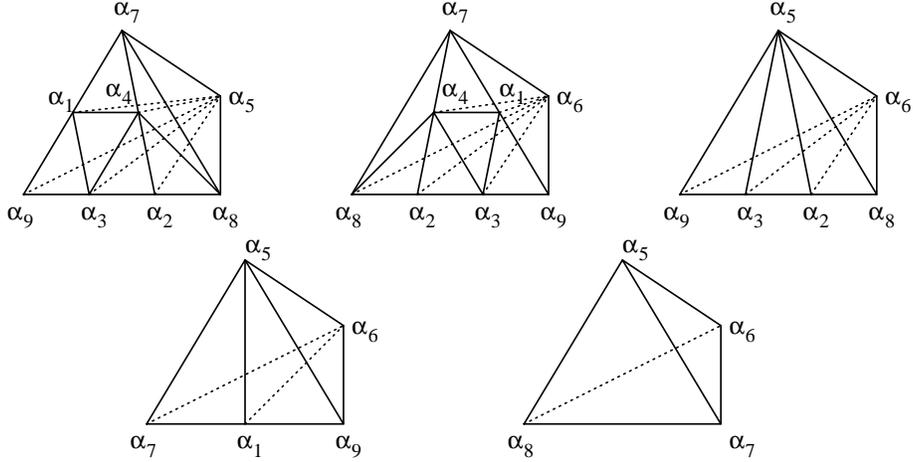
\centerline{Figure 13. The tetrahedra in resolution $\Delta_1$.}}$$
\endinsert
\fi

To obtain $Y$ as a mirror of $X$, we divide $X$ by the largest
``phase'' symmetry consistent with the trivial canonical bundle
condition \rGP. This is given by the following generators:
\eqn\eXpss{\eqalign{
 [z_0,z_1,z_2,z_3,z_4]&\to[\omega z_0,z_1,z_2,z_3,\omega^2z_4]\cr
 [z_0,z_1,z_2,z_3,z_4]&\to[z_0,\omega z_1,z_2,z_3,\omega^2z_4]\cr
 [z_0,z_1,z_2,z_3,z_4]&\to[z_0,z_1,\omega z_2,z_3,\omega^2z_4],\cr}}
where $\omega=\exp(2\pi i/3)$.
This produces a whole host of quotient singularities but since we are
only concerned with the complex structure of $Y$ we can ignore this
fact.

In light of the results of \rnewW\ and the discussion in section VI, we should
actually be more careful in our use of language here. To be more
precise, given the {\it Landau-Ginzburg\/} model $X_{\rm LG}$ whose
superpotential is specified in \eXfm, we can construct another
Landau-Ginzburg theory $Y_{\rm LG}$ as the orbifold of
$X_{\rm LG}$ by the group generated by \eXpss\ and having  the same
superpotential \eXfm. $Y_{\rm LG}$ is the mirror of
$X_{\rm LG}$. Up to now however, we had assumed that $X$ and $Y$
were smooth \CY\ manifolds. As shown in \rnewW\ and as will become
clear in section VI, the smooth \CY\ manifolds occupy a different
region of the same moduli space as the Landau-Ginzburg theory. Thus if
we deform {\it both\/} of our mirror pair $X_{\rm LG}$ and
$Y_{\rm LG}$, then we can obtain two smooth mirror manifolds $X$
and $Y$. If we wanted to compare all correlation functions of the
conformal field theories of $X$ and $Y$ then we would have to do this.
All we are going to do in this section however is to compare
information concerning the K\"ahler sector of $X$ with the complex
structure sector of $Y$. Information concerned with the complex
structure of $Y$ as a smooth manifold is isomorphic to that of
$Y_{\rm LG}$. Thus, there is no real need to deform $Y_{\rm LG}$
into a smooth \CY\ manifold. In figure 14 we show very roughly the
slice in which we do the calculation in this section. Note that this
figure is very oversimplified since the moduli space typically splits
into many more regions and indeed the whole point of this calculation
is to show that the area concerned spans more than one region.

\iffigs
\midinsert
$$\vbox{\centerline{\epsfxsize=10cm\epsfbox{catp-f14.ps}}
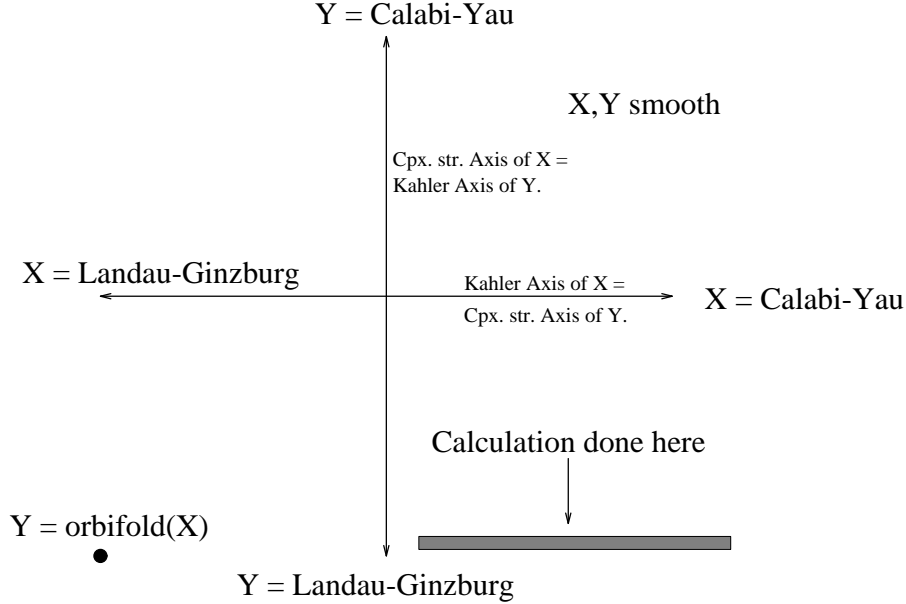
\centerline{Figure 14. Area of K\"ahler sector of $X$ and $Y$ where we
perform calculation.}}$$
\endinsert
\fi

The most general deformation of \eXfm\ consistent with this
$(\BZ_3)^3$ symmetry group is
\eqn\eComplex{\eqalign{W=a_0 z_0z_1z_2z_3z_4 +
a_1 z_2^3z_4^9 &+ a_2 z_3^6z_4^6 + a_3 z_3^3z_4^{12} +
a_4 z_2^3z_3^3z_4^3\cr
&+a_5z_0^3+a_6z_1^3+a_7z_2^6+a_8z_3^9+a_9z_4^{18} = 0.\cr}}
One can show that $h^{2,1}(Y)=7$. The group of reparametrizations of
\eComplex\ is indeed $(\IC^*)^5$ as required which shows that we
obtain 5 deformations of complex structure induced by deformations of
\eComplex. Note that for both $X$ and $Y$ we had 7 deformations of
which only 5 will be analyzed via toric geometry. It is no
coincidence that these numbers match --- it follows from the
monomial-divisor mirror map.

\subsec{The Moduli Spaces}

Let us now build the cones in $A_{n-1}(V)_\IR$ to form the partial
secondary fan. The method was outlined in subsection $3.8$. We first build
the $4\times9$ matrix $A$ with columns $\alpha_1,\ldots,\alpha_9$.
{}From this we build the $9\times5$ matrix $B$ whose columns span
the kernel of $A$.
The rows of $B$ give vectors in $A_{n-1}(V)_\IR$. Note that a change of
basis of the kernel of $A$ thus corresponds to a linear transformation
on $A_{n-1}(V)_\IR$. In order for us to translate the coordinates in
$A_{n-1}(V)_\IR$ into data concerning the coefficients $a_i$ in the
complex structure of $Y$ we need to chose a specific basis in
$A_{n-1}(V)_\IR$.

We have already fixed $a_0=1$. We still have a $(C^*)^4$ action on the
other $a_1,\ldots,a_9$ by which can fix 4 of these coefficients equal
to one. Let us choose $a_5=a_6=a_7=a_8=1$ and denote the matrix that
corresponds to this choice as $B_1$. Our 5 degrees of
freedom are given by $\{a_1,a_2,a_3,a_4,a_9\}$. We want that a
point with coordinates $(b_1,b_2,b_3,b_4,b_5)$ in $A_{n-1}(V)_\IR$
corresponds to $\{a_1=e^{2\pi i(c_1+ib_1)},a_2=e^{2\pi i(c_2+ib_2)},
\ldots,a_9=e^{2\pi i(c_5+ib_5)}\}$ for some value of the $B$-field
$(c_1,\ldots,c_5)$. This means our matrix $B_1$ should be of the form
$$B_1=\left(\matrix{1&0&0&0&0\cr0&1&0&0&0\cr0&0&1&0&0\cr0&0&0&1&0\cr
  B_{5,1}&B_{5,2}&B_{5,3}&B_{5,4}&B_{5,5}\cr
  B_{6,1}&B_{6,2}&B_{6,3}&B_{6,4}&B_{6,5}\cr
  B_{7,1}&B_{7,2}&B_{7,3}&B_{7,4}&B_{7,5}\cr
  B_{8,1}&B_{8,2}&B_{8,3}&B_{8,4}&B_{8,5}\cr0&0&0&0&1\cr}\right).
$$
$B_1$ is now completely
determined by the condition that its columns span the kernel of $A$.

For the actual calculation below, we use a slightly different set of
coordinates, choosing $\{a_0,a_1,a_2,a_3,a_4\}$ as the 5 degrees of freedom
and setting
$a_5=a_6=a_7=a_8=a_9=1$.
(We do this to express \eComplex\ in the
form: ``Fermat + perturbation'', in order
to more easily apply the calculational techniques of \rALR.)
The new basis can
be obtained from $B_1$ by using a $\IC^*$ action
$\lambda:z_4\to\lambda z_4$. We obtain the following matrix:
\def\ff#1#2{-{\scriptstyle #1\over #2}}
\def\vsttr{\vphantom{\displaystyle\sum}}
$$B=\left(\matrix{0&1&0&0&0\cr0&0&1&0&0\cr0&0&0&1&0\cr0&0&0&0&1\cr
  \vsttr\ff13&0&0&0&0\cr\vsttr\ff13&0&0&0&0\cr
  \vsttr\ff16&\ff12&0&0&\ff12\cr\vsttr\ff19&0&\ff23&\ff13&\ff13\cr
  \vsttr\ff1{18}&\ff12&\ff13&\ff23&\ff16\cr}\right).
$$

For each of the resolutions $\Delta_1,\ldots,\Delta_5$ we can now
construct the corresponding cone in $\Psf$ following the method in
subsection $3.8$ using the $B$ matrix above. These five cones are shown
schematically in figure 15 and the explicit coordinates in table 1.

\iffigs
\midinsert
$$\vbox{\centerline{\epsfxsize=10cm\epsfbox{catp-f15.ps}}
\centerline{Figure 15. The partial secondary fan.}}$$
\endinsert
\fi

\midinsert
{\ninepoint
\def\ilspace{\omit&height2.5pt&&&&\cr}
$$\vbox{\tabskip=0pt \offinterlineskip
\halign{\strut#&
\vrule#&\hfil\quad$#$\quad\hfil&\vrule#&\hfil\quad$#$\quad\hfil&
\vrule#\cr \tablerule\ilspace
&&v_1&&(-{1\over3},0,0,0,0)&\cr\ilspace
&&v_2&&(-{7\over{18}},-{1\over2},-{1\over3},-{2\over3},
    -{1\over6})&\cr\ilspace
&&v_3&&(-{1\over6},-{1\over2},0,0,-{1\over2})&\cr\ilspace
&&v_4&&(-{2\over9},0,-{1\over3},-{2\over3},-{2\over3})&\cr\ilspace
&&v_5&&(-{1\over9},0,-{2\over3},-{1\over3},-{1\over3})&\cr\ilspace
\tablerule\noalign{\vskip2pt}
\multispan{5}\hfil {\tenpoint Table 1a: Generators for first cone}\hfil\cr
}}
$$}
\endinsert

\midinsert
{\ninepoint
\def\ilspace{\omit&height2.5pt&&&&\cr}
$$
\def\ilspace{\omit&height1.5pt&&&&\cr}
\vbox{\tabskip=0pt \offinterlineskip
\halign{\strut#&
\vrule#&\hfil\quad#\quad\hfil&\vrule#&\hfil\quad#\quad\hfil&
\vrule#\cr \tablerule\ilspace
&&Resolution&&Generators&\cr\ilspace
\tablerule\ilspace
&&$\Delta_1$&&$v_1$, $v_2$, $v_3$, $v_4$, $v_5$&\cr\ilspace
&&$\Delta_2$&&$v_1$, $v_1-v_2+v_3+v_4$, $v_3$, $v_4$, $v_5$&\cr\ilspace
&&$\Delta_3$&&$v_1$, $v_2$, $v_2-v_3+v_4$, $v_4$, $v_5$&\cr\ilspace
&&$\Delta_4$&&$v_1$, $v_2$, $v_3$, $v_2+v_3-v_4+v_5$, $v_5$&\cr\ilspace
&&$\Delta_5$&&$v_1$, $v_2$, $v_3$, $v_2+v_3-v_4+v_5$, &\cr
\omit&&&&
\hphantom{$v_1$, $v_2$, }
$v_3+(v_2+v_3-v_4+v_5)-v_5$&\cr\ilspace
\tablerule\noalign{\vskip2pt}
\multispan{5}\hfil \tenpoint Table 1b: Generators for all cones\hfil\cr
}}$$}
\endinsert

Note that as expected the fan we generate, $\Psf$, is not a complete
fan and does not therefore correspond to a compact K\"ahler moduli space.

We now wish to translate this moduli space of K\"ahler forms into the
equivalent structure in the moduli space of complex structures on $Y$.
The way that we picked the basis in the $B$ matrix in this section
tells us exactly how to proceed.
For a point
$(u_0,u_1,\ldots u_4)$ in $A_{n-1}(V)_{\IR}$ we define
\eqn\eXek{
  w_k=e^{2\pi i(c_k+iu_k)}}
for any real $c_k$'s. This is then mapped to \eComplex\ by
\eqn\eXmm{\eqalign{a_i&=w_i^{-1},\quad i=0,\ldots,4\cr
  a_i&=1,\quad i=5,\ldots,9.\cr}}

\subsec{Results}

We have now done enough work to achieve one of our original goals,
i.e., given a specific topology of $X$ we can determine which
direction in the moduli space of $Y$ should be taken to find the
intersection numbers of $X$. All we do is is take a direction going
out to infinity within the interior of the corresponding cone in
$\Psf$ and convert this limit by \eXek.

\nref\rMOP{D. Markushevich, M. Olshanetsky and A. Perelomov, Commun.
Math. Phys. {\bf 111} (1987) 247.}

Just as was done in \rALR, it is easiest to compare intersection
numbers from the two methods by computing ratios which would be
invariant under rescaling of the monomials in \eComplex. There are
many such ratios although we will only consider 4 here for brevity. As
explained earlier, all intersection numbers are determined by the data in
$\Delta$ although we only outlined how to calculate the intersection
number of three homologically distinct cycles.
For the exact method of determining self-intersections we refer to
\rMOP\ or the method using the ``Stenley-Reisner ideal'' in \rOda.
Some ratios of
intersection numbers are given in table 2. In this table, $H$ refers
to the proper transform of the hyperplane in $\WCP{4}{6,6,3,2,1}$.
This can be identified with the point $O$ in $P^\circ$ and thus the
monomial with coefficient $a_0$. More will be said about this in the
next section.

\midinsert
\def\tablerule{\noalign{\hrule}}
\def\ilspace{\omit&height2pt&&&&&&&&&&&&\cr}
$$\vbox{\tabskip=0pt \offinterlineskip
\halign{\strut#&
\vrule#&\hfil\quad$#$\quad\hfil&\vrule#&\hfil\quad$#$\quad\hfil&
\vrule#&\hfil\quad$#$\quad\hfil&\vrule#&\hfil\quad$#$\quad\hfil&
\vrule#&\hfil\quad$#$\quad\hfil&\vrule#&\hfil\quad$#$\quad\hfil&
\vrule#\cr \tablerule\ilspace
&&\hbox{Resolution}&&\Delta_1&&\Delta_2&&\Delta_3
&&\Delta_4&&\Delta_5&\cr\ilspace\tablerule\ilspace
&&{(D_1^3)(D_4^3)\over(D_1^2D_4)(D_1D_4^2)}
&& -7 && 0/0 && 0/0 && \infty && 9 &\cr\ilspace
&&{(D_2^2D_4)(D_3^2D_4)\over(D_2D_3D_4)(D_2D_3D_4)}
&& 2 && 4 && 0 && 0/0 && 0/0 &\cr\ilspace
&&{(D_2D_3D_4)(HD_2^2)\over(D_2^2D_4)(HD_2D_3)}
&& 1 && 1 && 1 && 0 && 0/0 &\cr\ilspace
&&{(D_2D_3D_4)(HD_1^2)\over(D_1^2D_4)(HD_2D_3)}
&& 2 && 1 && \infty && 0/0 && 0 &\cr\ilspace
\tablerule
\noalign{\vskip2pt}
&\multispan{13}\hfil Table 2: Ratios of intersection numbers\hfil\cr
}}$$
\endinsert

\nref\rCandelas{P.~Candelas, Nucl. Phys. {\bf B298} (1988) 458.}
\nref\rVW{C. Vafa and N. Warner, Phys. Lett. {\bf 218B} (1989) 51.}

We now calculate the corresponding ratios for $Y$  choosing a direction in
the secondary fan space along the middle of the cone $\xi_\Delta$. The
value of the $B$ field is unimportant and we set it equal to zero.
Thus for a direction $(\lambda u_0,\ldots,\lambda u_4)$ we take the limit
\eqn\eXlim{
  a_i=\lim_{s\to\infty}s^{u_i},\quad\hbox{with}\;s=e^{2\pi\lambda}}
with $a_5,\ldots,a_9$ set equal to 1.
The $3$-point functions
are determined (up to an overall factor) by
using the simple structure of the
ring of chiral primary fields. This method was introduced in
\rCandelas. It can be shown from the $N=2$ superconformal algebra
\rVW\ that the monomials are members of the ring
\eqn\eCPring{\cpR={\IC[z_0,\ldots,z_4]\over({\partial W\over\partial z_0},
\ldots,{\partial W\over\partial z_4})},}
where the denominator on the right-hand side denotes the ideal
generated by the partial derivatives of $W$. If the hypersurface
generated by $W=0$ is smooth, i.e., if we are away from the
discriminant locus, then this ring structure determines the desired
couplings up to an overall factor.
The calculational method is that of \rALR; the results are shown in table
3. We use the symbol $\varphi_i$ to denote the field represented by the
monomial with coefficient $a_i$ (the factor $a_i$ is included).
As expected the results are in full agreement with the predictions in table 2
\foot{We emphasize that the predicted equalities result from our general
analysis of the previous sections and are not special to this illustrative
example. In fact, Batyrev
\ref\rBatQ{V. Batyrev, {\it Quantum cohomology rings of
toric manifolds}, MSRI preprint, 1993.}
 has recently generalized the  calculation
of \rAGM\ (which we have just reviewed) and shown that the predicted
equalities \rAGM\ are  true for the case of Calabi-Yau's and mirrors
realized as hypersurfaces in toric varieties.}.

\mark{B}\setbox\Bbox=\vbox{
{\ifx\answ\bigans\eightpoint\fi\hfuzz=10cm
\def\doots{\hbox to 4truept{$\dots$}}
\def\lquad{\hskip.3em\relax}
\def\tablerule{\noalign{\hrule}}
\def\ilspace{\omit&height2pt&&&&&&&&&&&&\cr}
$$\vbox{\tabskip=0pt \offinterlineskip
\halign{\strut#&
\vrule#&\hfil\quad$#$\quad\hfil&\vrule#&\hfil\lquad$#$\quad\hfil&
\vrule#&\hfil\lquad$#$\quad\hfil&\vrule#&\hfil\lquad$#$\quad\hfil&
\vrule#&\hfil\lquad$#$\quad\hfil&\vrule#&\hfil\lquad$#$\quad\hfil&
\vrule#\cr \tablerule\ilspace
&&\hbox{Resolution}&&\Delta_1&&\Delta_2&&\Delta_3&&
	\Delta_4&&\Delta_5&\cr\ilspace\tablerule\ilspace
&&\hbox{Direction}&&\omit\hfil$
({11\over9},1,{4\over3},{5\over3},{5\over3})
$\hfil&&\omit\hfil$
({7\over6},{1\over2},1,1,{5\over2})
$\hfil&&\omit\hfil$
({3\over2},{1\over2},2,3,{3\over2})
$\hfil&&\omit\hfil$
({13\over9},2,{5\over3},{4\over3},{4\over3})
$\hfil&&\omit\hfil$
({11\over6},{7\over2},1,1,{3\over2})
$\hfil&\cr\ilspace\tablerule\ilspace
&&{\langle \varphi_1^3\rangle\langle \varphi_4^3\rangle\over\langle
\varphi_1^2\varphi_4\rangle\langle \varphi_1\varphi_4^2\rangle}
&& -7-181s^{-1}+\doots &&  &&  && -{2\over5}s^2-{129\over250}s+\doots
&& 9+289s^{-1}+\doots &\cr\ilspace
&&{\langle \varphi_2^2\varphi_4\rangle\langle \varphi_3^2\varphi_4
\rangle\over\langle \varphi_2\varphi_3\varphi_4\rangle\langle
\varphi_2\varphi_3\varphi_4\rangle}
&& 2-5s^{-1}+\doots && 4-22s^{-1}+\doots && 0+2s^{-1}+\doots &&  &&
&\cr\ilspace
&&{\langle \varphi_2\varphi_3\varphi_4\rangle\langle \varphi_0\varphi_2^2
\rangle\over\langle \varphi_2^2\varphi_4\rangle\langle \varphi_0\varphi_2
\varphi_3\rangle}
&& 1+{1\over2}s^{-1}+\doots && 1+{3\over2}s^{-2}+\doots
&& 1+4s^{-1}+\doots && 0-2s^{-1}+\doots &&  &\cr\ilspace
&&{\langle \varphi_2\varphi_3\varphi_4\rangle\langle \varphi_0\varphi_1^2
\rangle\over\langle \varphi_1^2\varphi_4\rangle\langle \varphi_0\varphi_2
\varphi_3\rangle}
&& 2+27s^{-1}+\doots && 1-{1\over2}s^{-1}+\doots && -2s-33+\doots &&
&& 0+4s^{-2}+\doots &\cr
\tablerule
\noalign{\vskip2pt}
&\multispan{13}\hfil \tenpoint
Table 3: Asymptotic ratios of 3-point functions\hfil\cr
}}$$
}}
\ifx\answ\bigans\box\Bbox\fi

The monomial-divisor mirror map \rmdmm\ allows us to make more specific
comparisons. In general one might think that mirror symmetry gives
some correspondence in the large-radius limit only up to some factors. For
example, we might have
\eqn\eXmmm{D_2\sim \lambda_2a_2z_3^6z_4^6,}
and similar expressions for the other monomials and divisors, with
some constants
$\lambda_i$ being needed to specify the precise correspondence.
The invariant ratios calculated in table 3 are
designed so that the $\lambda_i$ factors cancel. The monomial-divisor
mirror map asserts however that $\lambda_i=1$. Thus we should only have one
overall scale factor undetermined. (This one scale factor comes from
the undetermined factor implicit in using the ring $\cpR$ to calculate
couplings.) As an example, in resolution $\Delta_1$ we have
intersection numbers:
\def\fr#1#2{{\textstyle{#1\over#2}}}
\eqn\eXino{\eqalign{D_1^3&=-3\cr D_4^3&=21\cr
	D_2^2D_4&=-6\cr D_3^2D_4&=-3.\cr}}
On $Y$, we fix the overall constant by demanding that
$\langle\varphi^3_0\rangle=1$. We then calculate
\eqn\eXinom{\eqalign{
  \langle\varphi_1^3\rangle &= -\fr1{486}(-3-\fr{195}2s^{-9}
	+\ldots)\cr
  \langle\varphi_4^3\rangle &= -\fr1{486}(21+\fr{183}2s^{-9}
	+\ldots)\cr
  \langle\varphi_2^2\varphi_4\rangle &= -\fr1{486}(-6-54s^{-9}
	+\ldots)\cr
  \langle\varphi_3^2\varphi_4\rangle &= -\fr1{486}(-3-\fr{45}2s^{-9}
	+\ldots)\cr
}}
This verifies the prediction made by the monomial-divisor mirror map
that, up to a single overall
factor, \eXinom\ should agree with \eXino\ to leading order.

\subsec{Discussion}

Let us recapitulate what we learn from this example. In subsection $4.2$
we presented a puzzle regarding the apparent asymmetry between complex
and K\"ahler moduli spaces.
We presented the two possible resolutions of this puzzle and a method
of adjudicating between them. In the context of the present example,
we have explicitly carried out this distinguishing test and shown that
the second resolution of subsection $4.2$ is verified. Namely, under mirror
symmetry   (part of) the complex structure moduli space of $Y$ is isomorphic to
the (partially)
{\it enlarged\/} K\"ahler moduli space of $X$ (and vice versa)
\foot{In the next section we will consider the fully enlarged
K\"ahler moduli space and show it to be isomorphic to the full moduli
space of complex structures on $Y$.}.
We have shown this by explicitly verifying that distinct points in the
complex structure moduli space of $Y$ are mapped to points in
{\it distinct regions\/} in the  (partially) enlarged K\"ahler moduli space of
$X$.

Although we have carried out this analysis in the context of a specific
example, we want to emphasize that our picture of the enlarged K\"ahler moduli
space, its toric geometric description and its relation to the complex
structure moduli space of the mirror, as presented in previous sections,
is completely general. The present example has served as an explicit
verification.

\newsec{The Fully Enlarged K\"ahler Moduli Space}

The analysis of the moduli spaces thus far leaves unanswered the
following interesting question.
As the fan $\Psf$ is not complete we can inquire as to what
happens if we take a limiting complex structure on $Y$
given by a direction not contained in $\Psf$. Up to this point, we
would be unable to give a corresponding mirror space $X$ for such a
limit. The reader might also have noticed the rather asymmetric way
we treated the point $O$ in $P^\circ$ and thus the monomial
corresponding to $a_0$. As we will see, a more symmetric treatment of
the polytope $P^\circ$ will open up the other regions of the K\"ahler
moduli space and yield a rich structure. It will turn out that these
other regions of moduli space will not have interpretations as smooth
\CY\ manifolds and are thus missed by classical $\sigma$-model ideas.
Thus this rich structure is one
example of the differences between classical and quantum geometry.

First let us reconsider the moduli space of complex structures on a
hypersurface within a weighted projective space. In \eaisone\ we
artificially singled out $a_0$ amongst the coefficients so that we
could reduce the $(\IC^*)^{n+1}$ action on the homogeneous
coordinates to the required $(\IC^*)^n$ action to obtain a toric
variety for the moduli space compatible with the moduli space of
K\"ahler forms.
As explained earlier, each monomial in \eCdd\ is represented by a point
in $P$ in the lattice $M$. Let us introduce a lattice
$M^+$ of
dimension $n+1$ obtained by adding one further generator to the
generators of $M$ in an orthogonal direction.
Thus $P$ lies in a hyperplane within $M^+_\IR$. It can be
seen that the $(\IC^*)^{n+1}$ action on any monomial is given by the
coordinates of this point in $M^+$. This now gives
rise to the following exact sequence
\eqn\ecsesp{
1\tto \exp(2\pi i\,M^{+*}_{\IC})\tto(\IC^*)^k\tto\MM_0\tto1.
}
The $(\IC^*)^k$ is the space of polynomials with non-zero coefficients
and $\MM_0$ is the resultant open subset of the
moduli space. This process is of course
entirely equivalent to \ecses. The difference now comes when we
compactify this space. The fact that we can now let $a_0\to0$ opens up
new possibilities reflected by the fact that some of the dimensions of
the vector spaces in the above exact sequence have increased.

Now let us reconsider the space of K\"ahler forms on a toric variety
specified by a fan $\Delta$, describing it in terms of the triangulation
of $P^\circ$ determined by $\Delta$.
Consider a simplex in $n$-dimensions with $n+1$ vertices. If we attach
a real number to each vertex then have specified an affine function on
the simplex whose value at each vertex is the number we specify.
Indeed the space of affine functions on this simplex can be given by
the space $\IR^{n+1}$ of numbers on the vertices. Consider now
attaching a real number to each point in the set $\AA=P^\circ\cap N$. This
can be taken to define a continuous (but not necessarily smooth)
function $\Psi_\Delta:{N_\IR}\to\IR$ which is a smooth
{\it affine\/} function in each
simplex. The space of all such functions is isomorphic to $\IR^r$, where
$r$ is the number of points in $P^\circ\cap N$.

We have previously described the K\"ahler forms on $V_\Delta$ in terms
of functions which are linear on each cone; we now modify this to
a description in terms of functions which are affine on each simplex.
The space of smooth linear functions on $N_\IR$ is the dual vector space
$N^*_\IR$. We can go to the space of smooth affine functions if we add
an extra dimension to get $N^{+*}_\IR$. We can then write down
the following exact sequence
\eqn\ecfes{
0\tto N^{+*}_\IR\tto\IR^r\tto A_{n-1}(V)_\IR\tto0,
}
where the space $\IR^r$ is interpreted as the space of functions $\Psi_\Delta$.
The K\"ahler forms are now specified by a convexity condition.  In fact,
we have the following analogue of \eKfms:
\eqn\eaffine{
\hbox{Space of K\"ahler forms on $V_\Delta$}\cong
{\hbox{Space of strictly convex $\Psi_\Delta$}\over\hbox{Space of smooth
$\Psi_\Delta$}}
}
where this time we take $\Psi_\Delta$ from the space $\IR^r$ of continuous
functions which are affine on each simplex.
of K\"ahler forms on $X_\Sigma$ will be a subspace of
$A_{n-1}(V)_\IR$.

We can now repeat the construction of cones in the partial secondary fan in a
slightly extended version of the algorithm in subsection $3.8$:
Extend the
coordinates of the $r$ points in $\AA$ to coordinates
in the $(n+1)$-dimensional space  $N^+_\IR$ simply by adding a
``1'' in the $(n+1)$th place and take $A^+$ to be the $(n+1)\times r$
matrix of these column vectors. Denote the columns of $A^+$ as $a_i, i
=1\ldots r$.
Let the kernel of $A^+$ be spanned by
the columns of the $n\times(r-n-1)$ matrix $B^+$. Let $b^+_i, i=1\ldots r$
be the resulting vectors in $A_{n+1}(V)_\IR$ whose coordinates are given
by the rows of $B^+$.

Now each maximal simplicial cone, $\sigma\in\Delta$, is specified by
$n+1$ elements of $\AA$, one of the elements being $O$. Thus when we
consider the complimentary set of $b_j$'s, we never include $b_0$.
Thus we build precisely the same cone $\xi_\Delta$ in the partial secondary
fan as we did in subsection $3.8$.

We know that this does not fill out the whole of the secondary fan.
It is known
however that if we consider the set of cones corresponding to all
possible triangulations based in $\AA$ consistent with the convexity of
$\Psi_{\Delta}$ then these do indeed
fill out the whole of $A_{n-1}(V)_\IR$ \rBFS. One should note that by all
possible triangulations, we include triangulations which may ignore
some of the points in $\AA$ (although not the points
on the vertices of $P^\circ$). The process of ignoring an interior
point and the way it relates to $\Psi_\Delta$ is shown in figure 16
(cp.~the flop shown in figure 11).
We will refer to removing a point in this way as a {\it
star-unsubdivision\/} (since the inverse process is often known as a
star-subdivision). Star unsubdivisions at an arbitrary point are not
necessarily allowed since the resultant network might not be a
triangulation but when they are allowed, they correspond to a
codimension 1 wall of two neighbouring cones in much the same way as a
flop did.

In general not all possible triangulations based in $\AA$ admit a strictly
convex function. A toric variety from such a triangulation does not
admit a K\"ahler metric \rOP. It turns out that in our
example there are no such triangulations. In more complicated examples
however one might have the situation where a smooth K\"ahler manifold
could be flopped into a smooth non-K\"ahler manifold. In such a case
the non-K\"ahler manifold would {\it not\/} be represented
in the space $A_{n-1}(V)_\IR$.

\iffigs
\midinsert
$$\vbox{\centerline{\epsfxsize=6cm\epsfbox{catp-f16.ps}}
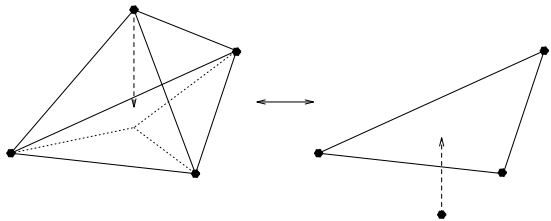
\centerline{Figure 16. A star-unsubdivision in terms of the function
$\Psi_\Delta$.}}$$
\endinsert
\fi

In our original construction the fan $\Delta$ corresponds to a toric
variety of complex dimension $n$ in which the \CY\ lives as a
hypersurface. A triangulation based in a set of points only determines a fan if
there is a point which is a vertex of every maximal-dimension simplex.
This point, which is the origin in $N$,
has coordinates $(0,0,\ldots,0,1)$ in $N^+$.
Since we now want to consider all triangulations we will have
cases where the triangulation does not determine
a fan. There is no obvious way of
associating a complex $n$-fold with many of the cones in the secondary
fan. As we are now working in $N^+$ however, there is an extra
dimension which offers the possibility of building a new fan
$\Delta^+$. This fan is shown in figure 17. Define the point $O^+$
in $N^+$ with coordinates $(0,0,\ldots,0,0)$.
Take every
simplex in the triangulation of $\AA$
and use it as the base of a cone with vertex
$O^+$. We now have all the necessary ingredients to build a toric
$(n+1)$-fold associated to {\it every\/} cone in the secondary fan.
Since all the points in $\AA$ lie in a hyperplane
at distance one from the origin, this $(n+1)$-fold
will have trivial canonical class. It will also be non-compact.

\iffigs
\midinsert
$$\vbox{\centerline{\epsfxsize=8cm\epsfbox{catp-f17.ps}}
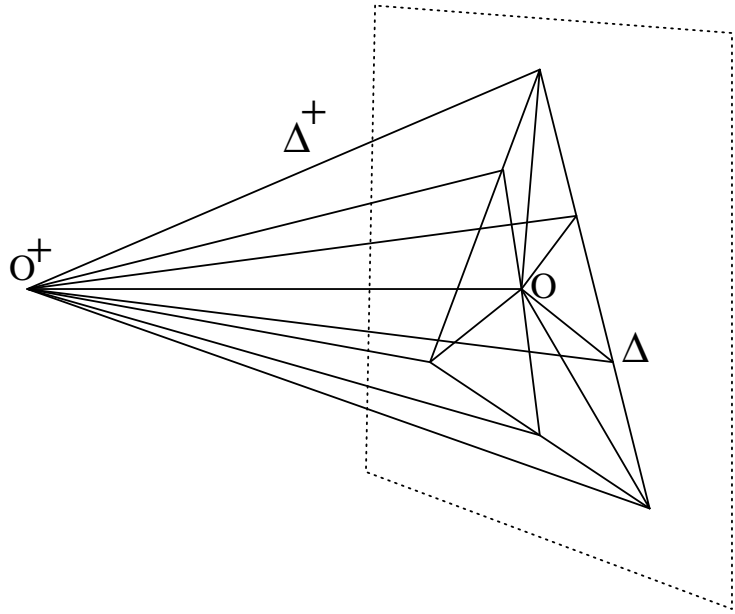
\centerline{Figure 17. The fan $\Delta^+$.}}$$
\endinsert
\fi

First take the example where $\Delta$ is a fan and so we have a toric
$n$-fold $V_\Delta$. In this case it can be shown that the toric
$(n+1)$-fold, $V_{\Delta^+}$, has a map
\eqn\elbnd{
\pi:V_{\Delta^+}\to V_\Delta,
}
such that our new space is simply a holomorphic
line bundle over $V_\Delta$.
Toric
geometry also tells us that $V_{\Delta^+}$ has vanishing first Chern
class. {\it It follows that $V_{\Delta^+}$ is the total space of
the canonical line bundle
of $V_\Delta$.}

At the other extreme we can take $\Delta\cap P^\circ$ to consist of one single
simplex, namely $P^\circ$. In this case $V_{\Delta^+}$ is the space
$\IC^{n+1}/G$ for some abelian group $G$. In our example we would
obtain $\IC^5/\BZ_{18}$. This is the target space for the
Landau-Ginzburg model. We have thus obtained the secondary fan version
of the argument in \rnewW, where it was shown that a
Calabi-Yau $\sigma$-model theory and a
Landau-Ginzburg theory occupy different regions of the
same moduli space.

We should be more specific about the above constructions for \CY\
manifolds and Landau-Ginzburg theories so that we will be able to
understand the more complicated examples that follow. The reader
should consult \rnewW\ for a more detailed discussion of these issues.
We have a
function $W:V_{\Delta^+}\to\IC$ (the superpotential) which is used to
define our theory.
This superpotential can be lifted to the space $\IC^{\AA}$ from which
$V_{\Delta^+}$ is obtained as a holomorphic quotient.
One also knows that $V_{\Delta^+}$ is
noncompact but will generally have compact
submanifolds algebraically embedded within it.
The general idea is that the space of vacua of the theory is the
critical point set of $W$ in
$\IC^{\AA}$ with the suitable quotient being taken.
If the superpotential $W$ is generic, the resulting
quotient is the intersection of the
compact part(s) of $V_{\Delta^+}$ with the locus
$W=0$. In the case of the regions studied in section V the compact
part is the smooth toric $n$-fold and the locus $W=0$ gives the \CY\ manifold
itself. In the case of the Landau-Ginzburg model the compact part is
simply the point at the origin and so is contained in $W=0$ anyway. We
also need to consider fluctuations about the vacuum. The potential for
such fluctuations is $|\del W|^2$. In the case of the \CY\ manifold,
smoothness tells us that fluctuations normal to the \CY\ manifold will
be massive (i.e., have quadratic potential) \rnewW. Thus, the target space is
precisely the \CY\ manifold. In the case of the Landau-Ginzburg theory
these fluctuations are massless and are analyzed in the usual way from
the superpotential $W$.

In addition to $\sigma$-models and Landau-Ginzburg theories,
there are many other possibilities.
In our example, the secondary
fan is divided up into 100 cones. That is, there are 100 different
triangulations with vertices in the point set $P^\circ\cap N$. Five of these
give smooth \CY\ manifolds and one
of these is the Landau-Ginzburg theory. There are 94 other regions
to explain!

Let us begin by analyzing the structure of the singularities of
$V_{\Delta^+}$. In our example, for every cone in the secondary fan
the corresponding $\Delta^+$ turns out to be simplicial.  Thus,
each cone $\sigma$ spanning $p$ dimensions in $\Delta^+$
is defined by the coordinates
of the $p$ vertices of the simplex $\sigma\cap N_\IR$ at the base of the cone.
The volume of that cone is then the volume of the polyhedron spanned
by $O^+$ and that simplex.  Volumes are normalized so that the unit
simplex in $\IR^p=\spn(\sigma)$ (with respect to the lattice
$N\cap \IR^p$) has volume 1.  (If we had normalized to make the unit
{\it cube\/} have volume 1, as is perhaps more common, then we would
have needed to multiply by $p!$ to obtain this ``simplicial'' volume.)
The cone
represents a smooth part of the toric variety iff this volume is 1.
That is, if the volume of a $p$-cone is greater than 1
then it represents a singularity along a $(n+1-p)$-dimensional
subvariety of $V_{\Delta^+}$. The fact that each cone is a cone over a
simplex tells us that the singularities in question are quotient
singularities. That is, a $p$-cone of volume $v$ represents a quotient
singularity locally of the form $\IC^p/G$ where $G$ has order $v$.
This gives the general statement about the correspondence between
Landau-Ginzburg models and \CY\ manifolds. The minimal triangulation based in
$P^\circ\cap N$ consisting of the single simplex $P^\circ$
is a target space of the form $\IC^{n+1}/G$ whereas a
maximal triangulation consisting of $|G|$ $(n+1)$-cones gives a smooth
manifold.

Consider now taking a triangulation corresponding to a smooth \CY\
manifold and star unsubdividing at a point such that $O$ is still a
vertex of each simplex of the triangulation. The $(n+1)$-dimensional
toric variety $V_{\Delta^+}$ can still be thought of as the canonical
sheaf of $V_{\Delta}$ but now $V_{\Delta}$ has acquired quotient
singularities. That is, $V_{\Delta}$ is an orbifold\foot{By
``orbifold'' we mean a space whose only singularities are locally quotient
singularities.}. This leads to a
rather striking conclusion as follows. Geometrically speaking,
the orbifolds that arise in the subject of superstring
compactification can just be thought of as special cases of \CY\
manifolds. Any quotient singularity arising from an abelian group
action in three
dimensions preserving the holomorphic 3-form can be
resolved to give a smooth \CY\ manifold. In this sense orbifolds appear to
live {\it within\/} the walls of the K\"ahler cone of a \CY\ manifold.
The reverse process to resolution, i.e., the {\it blowing-down\/} of a
divisor within the \CY\ manifold, appears as a path beginning in the
interior of the K\"ahler cone and ending on the boundary. In this
``classical''
setting \CY\ manifolds are more general than orbifolds in the sense
that they occupy a larger dimensional region of moduli space than do
orbifolds. In the light of our preceding discussion it appears that
within quantum algebraic geometry
one has to modify this viewpoint. Orbifolds now appear to occupy a
region of moduli space equal in dimension to that of the \CY\
manifold. One should think of orbifolds as not living within the
walls of the K\"ahler cone of a smooth \CY\ manifold but occupying
their own K\"ahler cone on the other side of the wall!

In our example 27 of the regions correspond to orbifolds. The
possible singularities encountered are
$\BZ_2$, $\BZ_3$, $\BZ_2\times\BZ_2$, $\BZ_4$ and
$\BZ_6$ quotients.

In these orbifold regions as well as in the smooth \CY\ region,
the space $V_{\Delta^+}$
may be thought of as a bundle
(in some generalized sense to include singularities) over a compact
space.  This also holds for Landau-Ginzburg theory.
In the case of the Landau-Ginzburg theory this compact space is a
point and in the \CY\ manifold and orbifold
cases this space is a toric $d$-fold. The other 67 regions of the
secondary fan correspond to spaces intermediate between these. The
dimension of the compact space may be determined as follows. Within
the any particular triangulation based in $P^\circ\cap N$
we may look for complete
fans. Any complete fan spanning $\IR^p$ corresponds to a toric
$p$-fold. Thus, the maximal $p$ for which a complete fan can be found
will give us the dimension of the complex space. The resultant space
is not necessarily irreducible however.

Figure 18 shows examples of spaces of intermediate dimension in the
case $d=2$.
In
example (a) there is no complete fan of dimension 2 in $N_\IR$ but there
is the fan as shown of dimension 1. Thus the dimension of the compact
part of the toric variety has gone down by 1 from the smooth \CY\
case. In figure (b) there is a dimension 2 fan (shown as $\Delta_2$)
but not all cones are elements of this fan. Another fan, $\Delta_1$ is
required to complete the triangulation. This should be interpreted as
follows: $\Delta_1$ represents a $1$-dimensional space and $\Delta_2$
represents a $2$-dimensional space. They intersect at a point since
they share one 2-dimensional cone. Example (a) is irreducible but (b)
is reducible into two irreducible components.

We are now in a position to associate a $(n+1)$-dimensional
toric variety with each of the 100 regions for our example. Each
5-dimensional cone in the secondary fan shares a codimension 1 wall
with another 5-dimensional cone. We will refer to these regions as
{\it neighbouring}. As it would be laborious to list a
detailed explanation of all 100 regions we will just look at the ones
neighbouring the examples we already know.

\iffigs
\midinsert
$$\vbox{\centerline{\epsfxsize=8cm\epsfbox{catp-f18.ps}}
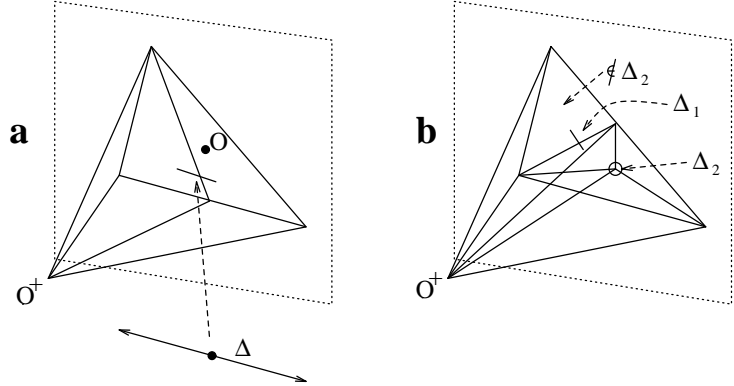
\centerline{Figure 18. Intermediate fans.}}$$
\endinsert
\fi

Consider first resolution $\Delta_1$. This has 3 flops to other smooth \CY's
and so this accounts for 3 of the 5 neighbours. Another possibility is
a star-unsubdivision on the point $\alpha_2$. This has the effect of
producing a curve of $\BZ_2$ quotient singularities within the \CY .
Star subdivisions on any other point do not give triangulations. The
fifth neighbour of resolution $\Delta_1$ is obtained from the triangulation
associated to
 resolution $\Delta_1$ in the following way. Remove all simplices
containing the point $\alpha_9$ (this removes four 4-plexes). Add in
the 4-plexes with vertices $\{\alpha_1,\alpha_3,\alpha_5,\alpha_6,
\alpha_9\}$ and
$\{\alpha_0,\alpha_1,\alpha_3,\alpha_5,\alpha_6\}$.
These simplices have volume 3 and 1
respectively. $V_{\Delta^+}$ in this case then refers to a variety
with a $\BZ_3$ quotient singularity at a point and it contains a
reducible subspace as the compact set. One part of this set is given
by the fan we still have around $O$ that corresponds to a 4-fold. This
fan does not cover $P^\circ$ however. The two 4-plexes that we added
form the fan of a $\IP^1$ touching the 4-fold at one point. The only
singularity of $V_{\Delta^+}$ lies on the $\IP^1$.

This resultant model has many interesting features. First note that we
have a $\BZ_3$ quotient singularity but it cannot be blown up and
maintain the trivial canonical sheaf condition. That is we cannot
perform a star subdivision of $\Delta^+$ on a point in $N$ to
obtain a smooth variety. Such a singularity is known as a terminal
singularity and only exists in 4 complex dimensions and higher. In
orbifold compactification one is accustomed to having twist fields associated
to quotient singularities which can be used as marginal operators. No
such field exists here.

In the case of the \CY\ model, intersecting the compact space with
$W=0$ lowered the dimension from 4 to 3. In contrast the condition
$W=0$ had no effect on the Landau-Ginzburg theory as its compact space
(i.e., the point at the origin)
was already contained in this locus. We should now address the
question of what happens to the $\IP^1$ which has appeared stuck onto
the side of the toric 4-fold. Will the $W=0$ condition intersect this
curve at a point or will it contain the whole curve? Figure 19 and
the following argument shows the latter is the case.

\iffigs
\midinsert
$$\vbox{\centerline{\epsfxsize=12cm\epsfbox{catp-f19.ps}}
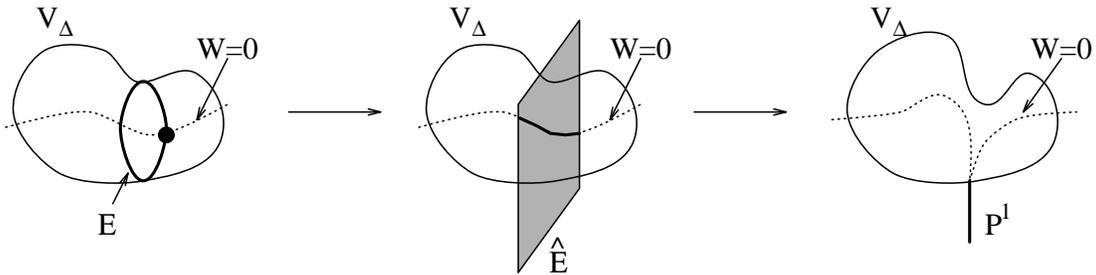
\centerline{Figure 19. An Exoflop.}}$$
\endinsert
\fi

The transition from resolution $\Delta_1$ to the model in question is similar
to a flop. The main difference is that in this new transition, the new
$\IP^1$ appears external to the \CY\ manifold. We will thus term it an
{\it exoflop}. The fan we described
above was obtained from the fan for the smooth \CY\ manifold by first
removing cones from the fan and then adding others in. This amounts to
blowing down along some point set in the original model and then
blowing up in a different way to obtain the new model. We could also
blow-up first and then blow-down afterwards to achieve the same
transition. In the case of a flop one blows up along a $\IP^1$ to
obtain $\IP^1\times\IP^1$ and then blows the other $\IP^1$ down to
obtain a $\IP^1$ not isomorphic to the original. In our example we
blow up $V_{\Delta^+}$ along
a 3-fold, $E$, which lives inside the compact 4-fold $V_\Delta$ for
resolution $\Delta_1$. This gives a 4-fold, $\hat E$, which looks locally
like $E\times\IP^1$. $\hat E$ is then blown-down to $\IP^1$. The $W=0$
locus intersects $E$ along a 2-fold and $\hat E$ along a 3-fold. Note
that if $p$ was contained in the $W=0$ locus in $E$ then $\IP^1\times
p$ is contained in the $W=0$ locus in $\hat E$. Thus, when we blow
down $\hat E$ to $\IP^1$, the whole of $\IP^1$ is contained in the
$W=0$ locus. Even though the 4-fold part of $V_\Delta$ does not acquire
any singularities during this process, the $W=0$ locus, and hence the
\CY\ space, does become singular as shown schematically in figure 19.

The \CY\ part of this model has no massless modes associated it except
in the tangent directions just as in the case of the more usual \CY\
manifold. There are deformations over the $\IP^1$ however which lead
to $|\del W|^2$ having higher than quadratic terms. Thus we have a kind of
Landau-Ginzburg type theory over this $\IP^1$ as in \rnewW.

Actually this is a generic feature of the 67 regions which are not
Landau-Ginzburg orbifolds, smooth \CY\ 3-folds or 3-dimensional
orbifolds. Namely, at least part of the target space has a kind of
Landau-Ginzburg fiber over it. These theories are referred to as
``hybrid'' models.

To sum up the results of the last few paragraphs the geometric
interpretation of the fifth neighbour of resolution $\Delta_1$ is as follows.
We have a reducible space with 2 irreducible components. One of these
components is a \CY\ space with a singularity where it touches the
other component. The other component is a
V-bundle (i.e., a bundle with quotient singularities in the fiber)
over a $\IP^1$ with
a Landau-Ginzburg type theory in the bundle. The $\CP1$ touches the
\CY\ space at one point.

The five neighbours of resolutions $\Delta_3$,
$\Delta_4$ and $\Delta_5$ are similar to the case
of resolution $\Delta_1$. The only possibilities are (a) flops, (b) blow-downs
to give
\CY\ spaces with quotient singularities, and (c) in each case one wall
corresponds to an exoflop. Resolution $\Delta_2$ has neighbours corresponding
to flops and blow-downs but the fifth wall is not an exoflop. In this
case the fifth possibility is a star-unsubdivision on the point $O$.
This results in there being no 4-dimensional complete fans within
$\Delta^+$. The structure of $\Delta^+$ is that of a 2-dimensional
complete fan consisting of 6 cones around the plane containing $\{
\alpha_4,\alpha_5,\alpha_6\}$. The resultant compact variety from this fan is
$\IP^2$ with 3 points blown up. This space is completely contained
within the $W=0$ locus and thus is the vacuum space for this theory.
There are massless modes around this vacuum so we obtain a
Landau-Ginzburg type theory in a V-bundle over a $\IP^2$ with 3 points
blown up.

We can generalize the concept of blowing down and its relation to
star-unsubdivisions for this example. As the point on which we have
unsubdivided the triangulation was the vertex of a complete fan in 4
dimensions, we have blown down a toric 4-fold. Hence, after
intersecting with $W=0$ we appear to have shrunk down the whole \CY\
manifold. There were exceptional divisors within the \CY\ manifold
however which had come from resolving quotient singularities. We did
not star-unsubdivide on these and so these retain their
dimension. Recall that these exceptional divisors corresponded to three
surfaces (treated as one for the purposes of toric geometry) contained
in the neighbourhood of the original 3 points corresponding to $\BZ_6$
quotient singularities, as well as three divisors generally of
the form $\IP^1\times$(curve) coming from the non-isolated $\BZ_2$
and $\BZ_3$ singularities. When we do the ``blow-down'' corresponding
to the star-unsubdivision at $\alpha_0$ we set to zero all distances within
the \CY\ manifold that did not arise from resolving singularities.
Thus, the three surfaces become one since they are now zero distance
apart and the three divisors of the form $\IP^1\times$(curve) become
$\IP^1$'s. These $\IP^1$'s are simply the 3 points blown up on the
$\IP^2$. This process is depicted in figure 20.

\iffigs
\midinsert
$$\vbox{\centerline{\epsfxsize=8cm\epsfbox{catp-f20.ps}}
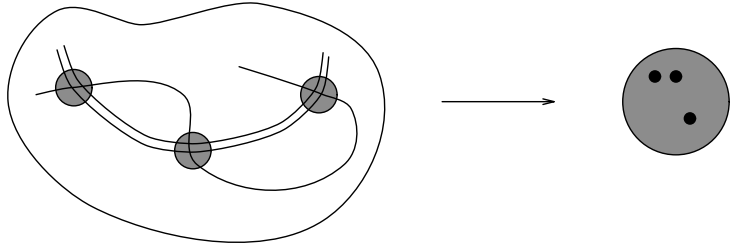
\centerline{Figure 20. Blowing down the whole \CY .}}$$
\endinsert
\fi

The above theory should be thought of as a natural step in the general
process of moving from a \CY\ manifold to a Landau-Ginzburg theory.
Whereas in the simpler case studied in \rnewW\ there was a direct
transition from a \CY\ 3-fold immediately down to a point giving a
Landau-Ginzburg theory, our example appears to proceed more naturally
by steps. The original \CY\ 3-fold moves down to a 2-fold and then one
can show there are models corresponding to 1-folds between this and
the Landau-Ginzburg point model.

The analysis of subsection $4.3$ can now be repeated for the whole
secondary fan. That is, for each cone we take a ``large radius limit''
by following a line out to $\infty$ within a cone. In the case of the
95 cones that do not give smooth \CY\ manifolds this will not, in
fact, be a purely large radius limit but rather some kind of generalized
geometric
limit. That is, it is the limit where the geometrical interpretation
becomes exact. If this limit is not taken then there are some kind of
instanton (or anti-instanton) corrections in the sense of \rnewW.
If a generic value of the $B$-field is taken, then transitions
from one cone to another have no singularities. Thus, transitions
between all the bizarre geometrical interpretations we have described
in this section are just as smooth as flop transitions in conformal
field theory.

\newsec{Conclusions}

A fascinating aspect of string theory is the interplay between
operations and constructs on the world-sheet and their corresponding
manifestations in spacetime. Familiar examples include the relationship
between world-sheet and spacetime supersymmetry and the connection between
world-sheet Ka\v c-Moody symmetries and spacetime gauge symmetries.
Part of the purpose of the present work has been to show another, rather
striking, correspondence. Namely, the world-sheet operation of deformation
by a truly marginal operator can, in certain circumstances, have a
macroscopic interpretation as a change in the topology of spacetime.
We have thus shown the truth of the long held suspicion that in string
theory there are  physical processes
leading to a change in the topology of spacetime. In fact, we have shown
that these processes are in no way special or exotic --- as they correspond
to changing the expectation value of a truly marginal operator, these
processes are amongst the most basic phenomena in conformal field theory.

More generally, the present work has set out to elucidate the
structure of \CY\ moduli space from a more global perspective
than has previously been undertaken. We have seen that such a vantage
point uncovers a wealth of rich and unexpected structure. The geometrical
interpretation of conformal field theory moduli space requires that we
augment the traditional K\"ahler moduli space associated with a
\CY\ space of a fixed topological type to the enlarged K\"ahler moduli.
The latter, as we have discussed, is a
space consisting of numerous regions adjoined along common walls. These
regions consist of the K\"ahler moduli spaces for: topologically distinct
\CY\ manifolds related by flops, singular orbifolds of smooth \CY\ manifolds,
Landau-Ginzburg conformal models and also the exotic ``hybrid'' conformal
field theories we have encountered. Moreover, we have shown that one can pass
in a physically smooth manner between these regions. In fact,
under mirror symmetry, the whole enlarged K\"ahler moduli space
corresponds to the (ordinary) complex structure moduli space of the mirror
\CY\ space. Thus, passing between regions in the K\"ahler moduli space
generically corresponds to nothing more than a smooth change in the
complex structure of the mirror manifold.

In addition to learning that spacetime topology change can be realized
in string theory, the need to pass to the enlarged K\"ahler moduli space
has resolved two important questions in mirror symmetry and shed new
light on the nature of orbifolds in string theory. For the former we
have seen that the work presented here resolves the apparent asymmetry
between complex and K\"ahler moduli spaces and also shows how the string
deals with singularities admitting distinct resolutions thereby clarifying
the global picture of mirror symmetry. For the latter we have seen that
notion of orbifolds as boundary points in the walls of the K\"ahler
cone of the \CY\ needs modification
in the context of conformal field theory/quantum algebraic
geometry. Rather,   orbifold theories
occupy their own regions in the
enlarged K\"ahler moduli space and  hence stand on equal footing with
smooth \CY\ manifolds.

We confined most of our  detailed discussion in section VI to those
regions in the enlarged K\"ahler moduli space corresponding to
well studied theories such as \CY\ and orbifold backgrounds. It is interesting
and important to gain an equally comprehensive understanding of
the regions corresponding to the less familiar hybrid theories.
In fact, if the example studied in section V is any indication, the
{\it majority\/} of moduli space corresponds to such theories!
Their investigation is therefore of obvious merit and we intend to
report on such studies shortly.

\bigbreak\bigskip\bigskip\centerline{{\bf Acknowledgements}}\nobreak
We thank R. Plesser for important
contributions at the inception and early stages of
this work.  We gratefully acknowledge numerous invaluable discussions
we have had with E.~Witten regarding both the present study and
his closely related work \rnewW.
We also thank S.~Katz, S.-S.~Roan, and R.~Wentworth for discussions.
B.R.G.\ thanks S.-T.~Yau for suggesting
the importance of studying flops in conformal field theory.
The work of P.S.A.\ was supported by DOE grant
DE-FG02-90ER40542, the work of B.R.G.\ was supported
by the Ambrose Monell Foundation, by a National Young Investigator award and
by the Alfred P. Sloan foundation,
and the work of D.R.M.\ was supported  by NSF grant DMS-9103827
and by an American Mathematical Society Centennial Fellowship.

\listrefs

\bye